\documentclass[prd,nofootinbib,twocolumn,superscriptaddress,floatfix]{revtex4-2}

\usepackage{graphicx}
\usepackage{amssymb,amsmath}
\usepackage{epstopdf}
\usepackage{aas_macros}
\usepackage{color}
\usepackage{multirow}
\usepackage{hyperref}
\usepackage[dvipsnames]{xcolor}

\newcommand{\lcdm}{{$\Lambda{\rm CDM}$}}

\newcommand{\impc}{{\rm Mpc$^{-1}$}}
\newcommand{\hmpc}{{\rm h Mpc$^{-1}$}}
\newcommand{\Ri}{{\mathcal{R}_i}}
\newcommand{\planck}{{\it Planck}}
\newcommand{\class}{{\sc CLASS}}

\begin{document}
\title{Enhanced Small-Scale Structure in the Cosmic Dark Ages}
\author{Derek Inman}
\thanks{derek.inman@ipmu.jp}
\affiliation{Kavli Institute for the Physics and Mathematics of the Universe (WPI), The University of Tokyo Institutes for Advanced Study, The University of Tokyo, Kashiwa, Chiba 277-8583, Japan}
\author{Kazunori Kohri}
\thanks{kohri@post.kek.jp}
\affiliation{Theory Center, IPNS, KEK, and Sokendai, 1-1, Oho, Tsukuba 305-0801, Japan}
\affiliation{Kavli Institute for the Physics and Mathematics of the Universe (WPI), The University of Tokyo Institutes for Advanced Study, The University of Tokyo, Kashiwa, Chiba 277-8583, Japan}
\date{\today}

\begin{abstract}
    We consider the consequences of a matter power spectrum which rises on small scales until eventually being cutoff by microphysical processes associated with the particle nature of dark matter.  Evolving the perturbations of a weakly interacting massive particle from before decoupling until deep in the nonlinear regime, we show that nonlinear structure can form abundantly at very high redshifts.  In such a scenario, dark matter annihilation is substantially increased after matter-radiation equality.  Furthermore, since the power spectrum can be increased over a broad range of scales, the first star forming halos may form earlier than usual as well.  The next challenge is determining how early Universe observations may constrain such enhanced dark matter perturbations.
\end{abstract}

\maketitle

\section{Introduction}

    The standard $\Lambda$CDM cosmological model consists of a cosmological constant dark energy ($\Lambda$), a cold and collisionless particle dark matter (CDM), as well as Gaussian initial perturbations that are both small and nearly scale invariant.  This model is precisely measured and well tested on large scales, both at early times through the cosmic microwave background (CMB) \citep{bib:FIRAS,bib:WMAP,bib:Planck2018,bib:ACT} and at late times through galaxy surveys \citep{bib:HSC,bib:DES,bib:eBOSS}.  On scales smaller than $\sim{\rm Mpc}$, however, it is much less constrained and its assumptions may not remain true \citep{bib:Bullock2017,bib:Allahverdi2021}.
    
    The default choice for CDM has been a weakly interacting massive particle (WIMP).  The WIMP particle decouples while non-relativistic and its thermal properties only impact clustering on $\sim{\rm pc}$ scales, setting a minimum halo mass comparable to that of the Earth \citep{bib:Green2004,bib:Green2005}.  However, it is now common, motivated by the non-observation of WIMPs in detector experiments \citep{bib:Arcadi2018} as well as potential discrepancies between numerical simulations of $\Lambda$CDM and observations of subhalos \citep{bib:Weinberg2015}, to consider other types of dark matter that can reduce clustering on larger scales.  Such particles could be dark matter with different interactions than the WIMP, leading to larger thermal velocities such as in warm dark matter \citep{bib:Bode2001} or interactions with other particles or itself leading to various types of damping \citep{bib:Buckley2014}.  Fuzzy dark matter is a different particle type of dark matter where quantum pressure actively opposes gravitational collapse \citep{bib:Hu2000,bib:Hui2017}.  While coming from different physical effects, these types of dark matter all have the qualitative effect of introducing a cutoff in the power spectrum below a certain scale.  The consequences of dark matter microphysics for structure formation has been parametrically studied in the ETHOS model \citep{bib:CyrRacine2016,bib:Vogelsberger2016} and such efforts are crucial in connecting observations to constraints \citep{bib:Bechtol2022,bib:Banerjee2022}.

    On the other hand, it is also interesting to consider scenarios that have more power on small scales.  An increase in small-scale power is a common prediction of many inflation models (e.g.~\citep{bib:Kohri2008,bib:Alabidi2012}) and has a number of interesting consequences.  If there is no cut-off to the primordial power spectrum and order unity superhorizon perturbations occur, then primordial black holes (PBH) can form when modes cross the horizon \citep{bib:Zeldovich1967,bib:Hawking1971,bib:Carr1974}.  Because order unity fluctuations are required, a complete non-observation of PBHs leads only to fairly weak constraints on the primordial power spectrum over a broad range of scales \citep{bib:Cole2018}.  Even if there is a mechanism to prevent PBH formation, the evolution of the Universe is still substantially altered by larger than expected perturbations.  In the radiation era, large perturbations in the cosmic plasma quickly lead to the formation of shocks, generating entropy and potentially gravitational waves \citep{bib:pen2016}.  Such shocks persist until density fluctuations are damped away by neutrino diffusion \citep{bib:jeong2014}.  This diffusion damping affects Big Bang Nucleosynthesis which leads to weak constraints on the primordial power on scales $k\sim 10^4-10^5 {\rm\ h^{-1}Mpc}$ \citep{bib:Inomata2016}.  On larger scales the power spectrum is constrained more strongly by spectral distortions arising due to Silk damping \citep{bib:Chluba2012}.

    If the dark matter also has enhanced perturbations, as expected for adiabatic initial conditions, then halos can begin to form soon after matter-radiation equality much earlier than the first $\Lambda$CDM halos.  Numerical studies of these early halo formation scenarios have been carried out \citep{bib:Gosenca2017,bib:Delos2018a} but focused on increased power on larger scales with heavier first halos.  It is also interesting to consider the potential interplay of a rising primordial power spectrum, called ``blue-tilted," that is eventually cutoff by damping processes associated with WIMP decoupling.  In this scenario, the first halos that form are still Earth mass, but exist at very high redshifts.  
    
    In standard $\Lambda$CDM, studying the first halos numerically is incredibly difficult due to the scale invariance of the power spectrum on small scales until the cutoff \citep{bib:Ishiyama2014,bib:Ishiyama2020}.  Since all small halos form at essentially the same time, the dynamic range required to reach low redshifts is impossible for a single N-body simulation.  Instead, some form of multi-resolution approach is required.  \citet{bib:Wang2020} used a set of 8 recursively nested zoom-in simulations to study the first halos until $z=0$.  \citet{bib:Takahashi2021} used 5 simulations with box sizes from $10{\rm\ Mpc}$ to $1{\rm\ kpc}$ to compute the nonlinear power spectrum at $z\ge10$.  With a blue-tilted power spectrum halos of different masses form at distinct redshifts and so we can study them at different times.
    
    The goal of this paper is to study how such a blue-tilted power spectrum may affect the cosmic dark ages, the times after recombination but before the first stars form.  In Section~\ref{sec:methods} we specify an explicit form of the primordial power spectrum consistent large-scale observations, compute the linear WIMP perturbations including effects of decoupling from the primordial plasma, and describe the setup of cosmological N-body simulations to analyze nonlinear structures.  In Section~\ref{sec:results} we report on the halos that form in these simulations, both the very light first halos as well as later ones that may host stars.  In Section~\ref{sec:discussion} we discuss potential consequences and observational constraints that may be impacted by the early formation of halos.  We conclude and discuss future directions in Section~\ref{sec:conclusion}.

\section{Methods}
\label{sec:methods}

Studying WIMP dynamics on small scales requires understanding their evolution from the very early Universe when they were still coupled to the cosmic radiation until the very late Universe where they have clustered into highly overdense halos.  In this section we attempt to calculate WIMP evolution including as many physical effects as possible.  One quantity of particular interest is the WIMP power spectrum, $\Delta^2_\chi(a,k)$.  If we assume perturbations start Gaussian, then WIMP perturbations are fully specified by $\Delta^2_\chi$ until nonlinear evolution occurs.  It is useful to separate the initial conditions, described by a primordial power spectrum $\Delta^2_\Ri(k)$, and the subsequent evolution, encoded in a transfer function $T_\chi(k,a)$, via $\Delta^2_\chi=T_\chi^2\Delta^2_\Ri$.  Once nonlinear evolution begins, a transfer function is no longer sufficient to specify the perturbations and instead one needs full N-body simulations.  We discuss our calculations of the primordial power spectrum, linear WIMP transfer functions and nonlinear N-body simulations in the following three subsections.

\subsection{Cosmological Parameters}
    \label{ssec:cosmoparam}
\begin{figure}
        \includegraphics[width=0.45\textwidth]{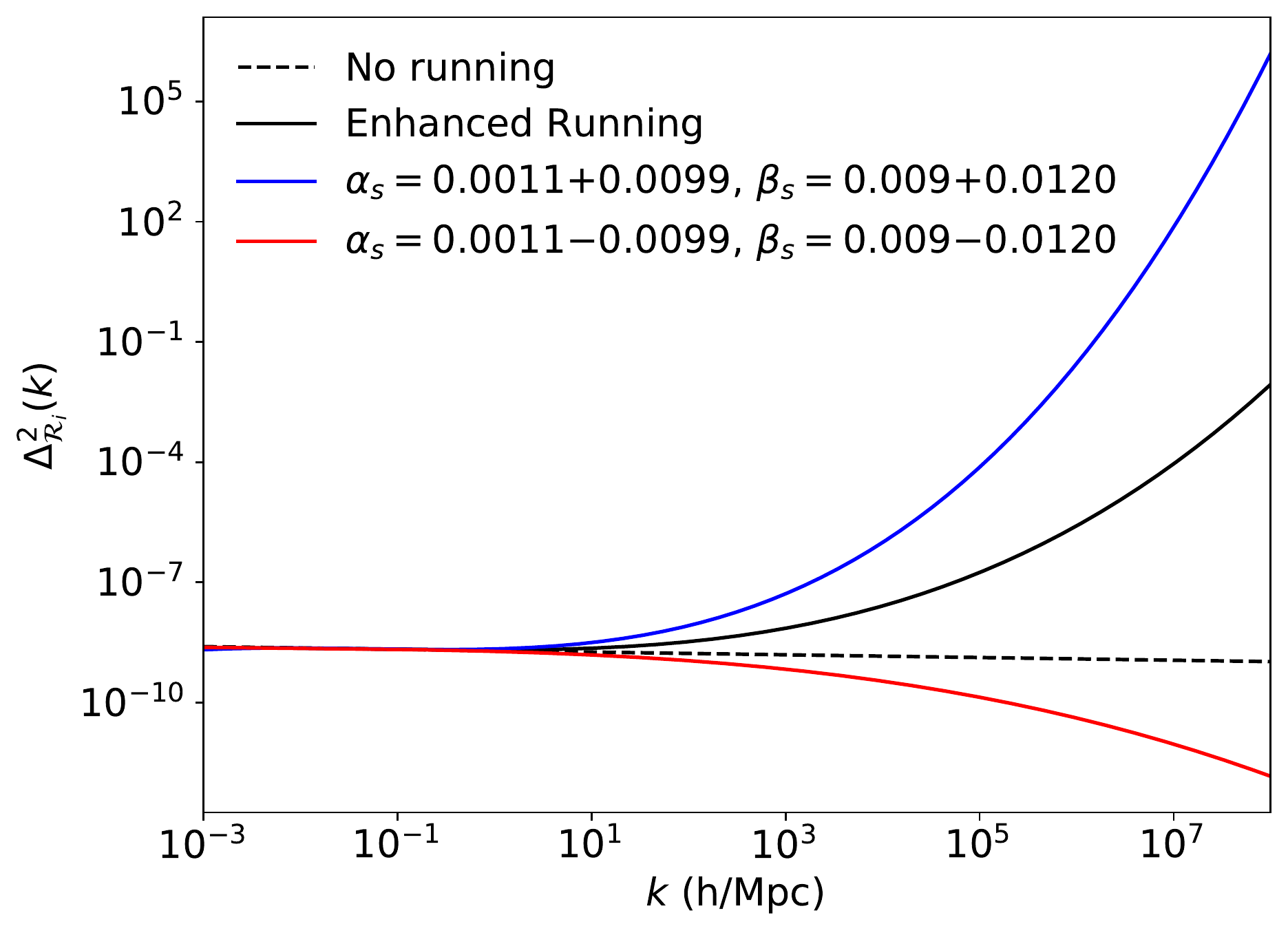}
        \caption{Initial curvature perturbation for various running and running-of-running parameters consistent on large scales with \planck{} measurements.  Simulations in this paper correspond to the `No Running' and `Enhanced Running' parameters.}
        \label{fig:ppk}
    \end{figure}

    Because $\Delta^2_\Ri$ is only weakly constrained on scales $k\gtrsim 1$ \hmpc, there is substantial freedom in what setup to consider.  A common extension (e.g.~\citep{bib:Kohri2008,bib:Alabidi2012,bib:Kohri2018}) of the regular parameterization with scalar amplitude $\mathcal{A}_s$ and spectral tilt $n_s$ is to include running ($\alpha_s$) and running-of-running ($\beta_s$) parameters:
    \begin{align}
        \Delta^2_\Ri=\mathcal{A}_s\left(\frac{k}{k_0}\right)^{n_s-1+\frac{\alpha_s}{2}\log\left(\frac{k}{k_0}\right)+\frac{\beta_s}{6}\log^2\left(\frac{k}{k_0}\right)}.
        \label{eq:delta2Ri}
    \end{align}
    The \planck{} experiment has precisely measured the values $\log\left(10^{10}\mathcal{A}_s\right)=3.043\pm0.014$ and $n_s=0.9647\pm0.0043$ for $k_0=0.05$ \impc, but the running parameters are consistent with zero: $\alpha_s=dn_s/d\log k=0.0011\pm0.0099$ and $\beta_s=d^2n_s/d^2\log k=0.009\pm0.012$ \citep{bib:Planck2018}. While there is no reason to apply Eq.~\ref{eq:delta2Ri} to arbitrarily small scales, it is {\it illustrative} to do so. In Fig.~\ref{fig:ppk} we show some examples of various power spectra consistent with \planck{} measurements on large scales but extrapolated to very small ones (noting that the extrapolation above unity is consistent with \planck{} but not as well motivated theoretically \citep{bib:Kohri2008}).  The difference with running parameters is substantial: while base \lcdm{} has a red tilted power spectrum, a strong blue tilt on small scales is perfectly possible.
    
    For the purposes of this work, we select cosmological parameters that are broadly consistent with \planck{} \citep{bib:Planck2018}: $\mathcal{A}_s=2.15\times10^{-9}$, $n_s=0.966$, $h=0.675$, $\Omega_c=0.26$, $\Omega_b=0.05$ and $z_{\rm eq}=3374$.  We assume a flat Universe and so $\Omega_\Lambda=1-\Omega_m-\Omega_r$ with $\Omega_m=\Omega_c+\Omega_b$ and $\Omega_r=\Omega_m/(1+z_{\rm eq})$.  We will contrast two choices of running parameters: a `No Running' scenario with $\alpha_s=\beta_s=0$ and an `Enhanced Running' one with $\alpha_s=0.002$ and $\beta_s=0.01$.  To describe WIMPs we require three parameters: their mass ($m_\chi$), decoupling temperature ($T_d$) and how their momentum transfer rate with standard model particles depends on temperature $\gamma\propto T^{2+n_\gamma}$.  Equivalently this can be thought of as when they decouple (at scalefactor $a_d$ or conformal time $\eta_d$), how fast the decoupling occurs ($n_\gamma$) and how warm the resulting dark matter is ($\propto T_d/m_\chi$).  In this work, we will consider a WIMP with mass $m_\chi=100$ GeV, decoupling at a temperature $T_d=10$ MeV and with $\gamma\propto T^6$ ($n_\gamma=4$), similar to \citep{bib:Bertschinger2006}.

\subsection{Transfer Functions}
    WIMPs undergo a number of decoupling processes in the early radiation era.  The first is chemical decoupling, or ``freeze-out," and occurs when the radiation temperature drops below the WIMP mass, $m_\chi\sim\mathcal{O}({\rm 100  GeV})$ \citep{bib:Bringmann2009}.  After this time, WIMP particles are no longer produced and their comoving number density is constant.  However, they can continue to scatter with the plasma until much lower temperatures,  $T_d\sim\mathcal{O}({\rm 10 MeV})$, and so remain thermally and kinetically coupled to the radiation fluid.  As the Universe continues to cool, all scattering with the standard model stops and the WIMPs become dark matter.  Interestingly, at this time they also stop interacting gravitationally with the standard model on subhorizon scales \citep{bib:Voruz2014}.  This is because the photon fluid is quickly oscillating rapidly relative to cold dark matter, and so gravitational accelerations sourced by radiation perturbations average to zero.  This continues until recombination when the baryons decouple as well.
    
    On scales that cross the horizon after WIMP decoupling, neutrino decoupling and electron-positron annihilation (all of which occur around $\mathcal{O}({\rm MeV})$ temperatures or $k\lesssim 10^4$ \hmpc), WIMPs behave like CDM and highly accurate transfer functions can be obtained from numerical Boltzmann codes like \class{} \citep{bib:class2011} or analytic subhorizon approximations like the one provided in \citet{bib:Hu1996}.  However, we are particularly interested in smaller scales as it is here that $\Delta_\chi^2$ peaks. A simple approach is to simply take the pure CDM transfer function and introduce a cutoff at a scale characteristic of WIMPs, $k\sim10^6$ \hmpc.  However, this misses the acoustic oscillations imprinted onto the WIMP transfer functions as they decouple.  To include such acoustic oscillations one can solve fluid equations including a kinetic coupling term with the photons \citep{bib:Loeb2005}. However, to fully describe the WIMPs from before decoupling until today it is necessary to solve the Boltzmann equation coupled to the Einstein field equations \citep{bib:Bertschinger2006}.  In this section we first find a general integral equation for the Boltzmann equation.  Because it is difficult to evaluate in generality, we then find an approximate solution by first solving the equations around decoupling when exact solutions of the Einstein field equations can be used, and then propagating the perturbations forwards using a collisionless approximation \citep{bib:Loeb2005,bib:Bertschinger2006}.
    We normalize transfer functions such that the superhorizon curvature perturbation $\Ri$ is unity \citep{bib:Lesgourgues2013} and use standard cosmological initial conditions \citep{bib:Ma1995}.
    
    \subsubsection{Boltzmann-Fokker-Planck Equation}
        For a WIMP scattering with relativistic particles the Boltzmann equation is given by \citep{bib:Bertschinger2006,bib:Binder2016}:
        \begin{align}
            \dot{f}+\frac{\vec{v}}{a}&\cdot\vec{\nabla}_x f + \left[ \vec{v}\dot{\phi}-a\vec{\nabla}_x\psi\right]\cdot\vec{\nabla}_vf\nonumber\\
            &=a\gamma(1+\psi)\vec{\nabla}_v\cdot\left[ (\vec{v}-a\vec{V}_R)f+\frac{a^2T_R}{m_\chi} \vec{\nabla}_v f\right]
            \label{eq:bfp}
        \end{align}
        where $f$ is the WIMP phase space density, $\dot{f}=\partial f/\partial\eta$ where $\eta$ is the conformal time, $\vec{\nabla}_x$ is the positional gradient in comoving coordinates, $\vec{v}$ is the comoving particle momentum $\vec{q}$ divided by its mass $m_\chi$, $\vec{\nabla}_v$ is a gradient with respect to velocity, $\vec{V}_R$ and $T_R$ are the plasma velocity and temperature, $\gamma$ is the momentum transfer rate with the plasma, and $\phi$ and $\psi$ are the scalar potentials in the notation of \citet{bib:Ma1995}.  The right hand side of Eq.~\ref{eq:bfp} is the Fokker-Planck collision operator which is appropriate when the momentum change per scattering event is small, although it can also be accurately used in more general contexts by matching the drift and diffusion terms to the collision operator \citep{bib:AliHaimoud2019,bib:Gandhi2022}.  We linearize this equation by taking $f=f_0+f_1$, $T_R=T_0(1+\delta_T)$, and $\gamma=\gamma_0(1+\delta_\gamma)$ with $\vec{V}_R$, $\phi$ and $\psi$ always first order.  We obtain the zeroth order equation
        \begin{align}
            \dot{f}_0-a\gamma_0\vec{\nabla}_v\cdot\left[\vec{v}f_0+\frac{a^2T_0}{m_\chi}\vec{\nabla}_vf_0\right]=0
            \label{eq:bfp0}
        \end{align}
        and first order equation:
        \begin{align}
            \dot{f}_1+\frac{\vec{v}}{a}\cdot\vec{\nabla}_xf_1 -a\gamma_0\vec{\nabla}_v\cdot\left[\vec{v}f_1+\frac{a^2T_0}{m_\chi}\vec{\nabla}_vf_1\right]=S
            \label{eq:bfp1}
        \end{align}
        where the source term is given by:
        \begin{widetext}
        \begin{align}
            S(\eta,\vec{x},\vec{v})=a\gamma_0\vec{\nabla}_v\cdot\left[(\delta_\gamma+\psi)\left(\vec{v}f_0+\frac{a^2T_0}{m_\chi}\vec{\nabla}_vf_0\right) -  a\vec{V}_Rf_0+\frac{a^2T_0}{m_\chi}\delta_T\vec{\nabla}_vf_0\right] -\left[\vec{v}\dot{\phi}-a\vec{\nabla}_x\psi\right]\cdot\vec{\nabla}_vf_0
            \label{eq:srcv}
        \end{align}
        \end{widetext}
        and is not directly dependent on $f_1$ but can depend on its moments through $\phi$ and $\psi$. The mean density is normalized to unity $\int d^3\vec{v} f_0=1$ and the WIMP density perturbation is given by $\delta_\chi=\int d^3\vec{v} f_1$.
        
        The Boltzmann-Fokker-Planck equation is very challenging to solve.  In non-cosmological contexts, \citet{bib:Chandrasekhar1943} and \citet{bib:Dougherty1964} obtain solutions along characteristics; however, the cosmological WIMP equation has many more time dependent quantities than the ones considered there. While \citet{bib:Bertschinger2006} numerically solved Eq.~\ref{eq:bfp1} using an eigenfunction approach, computing the full phase space is quite excessive if one is only interested in lower moments like the density and velocity.  These moments satisfy a set of differential equations that can be evolved numerically \citep{bib:CyrRacine2016,bib:Kamada2017}, however this introduces a different problem: the moment equations suffer from the well known lack of closure as each equation contains a new moment and so an infinite number of equations are required \citep{bib:Uhlemann2018}.  We therefore take a different approach and first convert Eq.~\ref{eq:bfp1} into an integral equation.  When formulated as integral equations, the equation for the density contrast decouples from higher moments, similar to the case of collisionless dynamics (e.g.~\citep{bib:Gilbert1966,bib:Bond1983,bib:AliHaimoud2013,bib:Ji2022}).  Higher moments can then be computed either through their own integral equations or by solving a finite set of differential equations.
        
        To obtain integral solutions to Eqs.~\ref{eq:bfp0} and \ref{eq:bfp1}, we first perform a Fourier transform in both position and velocity space:
        \begin{align}
            f(\eta,\vec{k},\vec{h})=\int d^3\vec{v}d^3\vec{x}e^{-i\vec{k}\cdot\vec{x}-i\vec{h}\cdot\vec{v}}f(t,\vec{x},\vec{v}).
        \end{align}
        $f(\eta,\vec{k},\vec{h})$ is called the moment generating function as we can obtain moments by taking gradients in $\vec{h}$ and then setting $\vec{h}=0$, in particular $\delta_\chi=f(\eta,\vec{k},0)$ \citep{bib:Uhlemann2018}.  In the context of the Boltzmann-Fokker-Planck equation, the velocity Fourier transform has been used to study the (linear) dynamics of collisional plasmas \citep{bib:Lenard1958,bib:Karpman1967,bib:Catto1979}.
        After making this transformation, Eq.~\ref{eq:bfp0} becomes:
        \begin{align}
            \dot{f}_0+a\gamma_0\vec{h}\cdot\vec{\nabla}_hf_0+a\gamma_0\frac{a^2T_0}{m_\chi}h^2f_0=0.
        \end{align}
        The characteristic equation is $d\vec{h}/d\eta'=a\gamma_0\vec{h}$ which has the solution for $0\le\eta'\le\eta$:
        \begin{align}
            \vec{h}_\eta(\eta')=\vec{h}(\eta)\exp\left[-\int_{\eta'}^\eta a\gamma_0 d\eta''\right].
        \end{align}
        where we use the notation that $_\eta$ indicates that a function is parameterized via the final conformal time $\eta$ instead of an initial conformal time, $\eta\rightarrow0$.

        We can solve the background equation along the characteristics via an integrating factor:
        \begin{align}
            f_0(\eta)=f_0(\eta\rightarrow0)\exp\left[-\int_{0}^{\eta} a\gamma_0\frac{a^2T_0}{m_\chi} h_\eta^2(\eta') d\eta' \right].
        \end{align}
        Taking cold initial conditions consistent with tight coupling, i.e.~with  $f_0(\eta\rightarrow0)=1$, the solution is at all times a Maxwell-Boltzmann distribution:
        \begin{align}
            f_0(\eta,v)=\frac{1}{(2\pi\sigma^2)^{3/2}}\exp\left[-\frac{1}{2}\frac{v^2}{\sigma^2}\right]& \nonumber \\
            \therefore f_0(\eta,h)=\exp\left[-\frac{1}{2}h^2\sigma^2\right]&
        \end{align}
        with velocity dispersion:
        \begin{align}
            \sigma^2(\eta)=2\int_0^\eta d\eta' a\gamma_0\frac{a^2T_0}{m_\chi}\exp\left[-2\int_{\eta'}^\eta a\gamma_0 d\eta''\right].
            \label{eq:sigma2}
        \end{align}
        
        Eq.~\ref{eq:bfp1} can be solved in the same way, although it is of course more complicated due to the advective and source terms.  After Fourier transforming it becomes:
        \begin{align}
            \dot{f}_1-\frac{\vec{k}}{a}\cdot\vec{\nabla}_hf_1+a\gamma_0\vec{h}\cdot\vec{\nabla}_hf_1&+a\gamma_0a^2\frac{T_0}{m_\chi}h^2f_1\nonumber\\&=S(\eta,h(\eta))
        \end{align}
        where we omit $k$ dependence for notational simplicity.
        The characteristic equation is $d\vec{h}/d\eta'=a\gamma_0\vec{h}-\vec{k}/a$ which has the solution:
        \begin{align}
            \vec{h}_\eta(\eta')=\vec{h}(\eta)\exp\left[-\int_{\eta'}^\eta a\gamma_0d\eta''\right] + 
            \vec{k}\frac{\eta_d}{a_d} u_\eta(\eta')
            \label{eq:hee}
        \end{align}
        where $\vec{h}_\eta(\eta')=(\vec{k}\eta_d/a_d) u_\eta(\eta')$ is the solution to characteristics terminating at $\vec{h}(\eta)=0$, and the dimensionless $u_\eta(\eta')$ is:
        \begin{align}
            u_\eta(\eta')=\int_{\eta'}^\eta \frac{d\eta''}{\eta_d}\frac{a_d}{a}\exp\left[-\int_{\eta'}^{\eta''}a\gamma_0d\eta'''\right].
            \label{eq:uee}
        \end{align}
        The perturbations can be solved along the characteristics via an integrating factor yielding the moment generating function:
        \begin{align}
            &f_1(\eta)=f_1(\eta\rightarrow0)\exp\left[-\int_{0}^{\eta}a\gamma_0\frac{a^2T_0}{m_\chi}h^2_\eta(\eta')d\eta'\right] 
            \nonumber\\&+\int_{0}^{\eta}d\eta' S(\eta',\vec{h}_\eta(\eta'))\exp\left[-\int_{\eta'}^{\eta}a\gamma_0\frac{a^2T_0}{m_\chi} h^2_\eta(\eta'')d\eta''\right].
            \label{eq:f1}
        \end{align}
        Initial conditions can be set by Fourier transforming the tight coupling solution \citep[Eq.~31]{bib:Bertschinger2006}:
        \begin{align}
            f_1(\eta\ll\eta_d,\vec{h})=\left[\delta_\chi+i\vec{h}\cdot a\vec{V}_\chi-\frac{1}{2}h^2\sigma^2\delta_{T_\chi}\right] f_0(\eta,h)
        \end{align}
        where $\vec{V}_\chi$ and $\delta_{T_\chi}$ are the WIMP velocity and temperature perturbations.  For our case with $f_0(\eta\rightarrow0)=1$ and no initial superhorizon velocity, this is just $f_1(\eta\rightarrow0)=\delta_\chi(\eta\rightarrow0)$.
        Fortunately, we do not need to perform the inverse Fourier transform of Eq.~\ref{eq:f1}: to obtain the density contrast we can simply set $\vec{h}(\eta)=0$ in Eq.~\ref{eq:hee} \citep{bib:Catto1979}:
        \begin{align}
            &\delta_\chi(\eta)=\delta_\chi(\eta\rightarrow0)G_\eta(\eta\rightarrow0) + \int_0^\eta d\eta' S_\eta(\eta')G_\eta(\eta')\label{eq:delta_x}
        \end{align}
        where we have separated the diffusion damping factors and source perturbations:
        \begin{widetext}
        \begin{align}
            G_\eta(\eta')&=\exp\left[-\frac{1}{2}\frac{T_d}{m_\chi}(k\eta_d)^2\left(\frac{\sigma^2}{\sigma_d^2}u^2_\eta(\eta')+2\int_{\eta'}^\eta a\gamma_0\frac{a^2}{a_d^2}\frac{T_0}{T_d} u_\eta^2(\eta'')d\eta''\right)\right] \label{eq:Gee}\\
            S_\eta(\eta')&=3\dot{\phi}-u_\eta(\eta')\left[a\gamma_0\frac{a}{a_d}(\eta_d\theta_R)+\frac{a}{a_d}k^2\eta_d\psi\right] \nonumber\\
            &+\frac{T_d}{m_\chi}(k\eta_d)^2u^2_\eta(\eta')\left[a\gamma_0\left(\frac{\sigma^2}{\sigma_d^2}-\frac{a^2}{a_d^2}\frac{T_0}{T_d}\right)\left(\delta_\gamma+\psi\right)-a\gamma_0\frac{a^2}{a_d^2}\frac{T_0}{T_d}\delta_T-\frac{\sigma^2}{\sigma_d^2}\dot{\phi}\right]\label{eq:src}
        \end{align}
        \end{widetext}
        where $\theta_R=i\vec{k}\cdot\vec{V}_R$ is the velocity divergence and $\sigma_d^2=a_d^2T_d/m_\chi$.
        
        Examining the Gaussian damping in Eq.~\ref{eq:Gee}, we see that there are two distinct cutoffs associated with diffusion, corresponding to $\eta'\ll\eta_d$ and $\eta'\gg\eta_d$.  The first is an integrated scale arising from the diffusive term in Eq.~\ref{eq:f1} with a characteristic damping wavenumber $k_{D}$:
        \begin{align}
            \left(\frac{T_d}{m_\chi}(k_{D}\eta_d)^2\right)^{-1}=2\int_{\eta'}^\eta a\gamma_0\frac{a^2}{a_d^2}\frac{T_0}{T_d} u_\eta^2(\eta'')d\eta''.
            \label{eq:kDD}
        \end{align}
        and for WIMPs we can take $\eta'\sim0$ in the integral to obtain the minimum $k_D$.  The second cutoff, which we denote $k_{S}$, is associated with dynamical streaming of particles and arises due to $f_0$ being a Gaussian in Eq.~\ref{eq:srcv}:
        \begin{align}
            \left(\frac{T_d}{m_\chi}(k_{S}\eta_d)^2\right)^{-1}=\left(a_d^2\frac{T_d}{m_\chi}\right)^{-1}\sigma^2(\eta')u_\eta^2(\eta').
            \label{eq:kFS}
        \end{align}
        Collisionless damping occurs after decoupling, $\eta'\gtrsim\eta_d$ and so this gives the free-streaming scale.  In addition to these diffusion cutoff scales, there is also frictional damping arising even for $T_d/m_\chi=0$ \citep{bib:Bertschinger2006}.  Estimates of this form of damping can be obtained via the steepest descent approximation \citep{bib:Bertschinger2006,bib:Kamada2018}.
        
        \subsubsection{Decoupling Solution}
        \label{ssec:decsoln}
        
        We now consider times around WIMP decoupling, which we will assume occurs deep in the radiation era (with Hubble rate $H=H_r/a^2$ and scalefactor $a=H_r\eta$) and during a period of constant entropy density ($T_0\propto 1/a$).  The decoupling parameters mentioned in the previous section can therefore be related by $a_d=H_r\eta_d$, $T_0/T_d=a_d/a$, $H_d=H_r/a_d^2$ and $a_dH_d=\eta_d^{-1}$.  We generally intend to consider a momentum transfer rate $\gamma\propto T^6$; however, other power laws are possible, e.g.~those in \citep{bib:Kamada2018,bib:Stadler2019}, and so we solve the more general $T^{2+n_\gamma}$ with $n_\gamma>0$ required for decoupling (it is also possible for the rate to be much more complicated, such as for charged massive particles \citep{bib:Kamada2017}).  For this more general momentum transfer rate, the linearized scattering rate can be parameterized as:
        \begin{align}
            \gamma=\frac{n_\gamma}{2}H_d\left(\frac{T_0}{T_d}\right)^{2+n_\gamma}(1+\delta_\gamma)
        \end{align}
        with $\delta_\gamma=(2+n_\gamma)\delta_T=(2+n_\gamma)\delta_R/4$.  Using the radiation background expansion, we furthermore have:
        \begin{align}
            a\eta\gamma_0=\frac{n_\gamma}{2}\frac{1}{y^{n_\gamma}}
        \end{align}
        where $y=a/a_d=\eta/\eta_d=T_d/T_0$.  Eq.~\ref{eq:sigma2} can be integrated analytically yielding:
        \begin{align}
            \sigma^2(y)=\sigma^2_d\exp\left[\frac{1}{y^{n_\gamma}}\right]\Gamma\left[\frac{n_\gamma-1}{n_\gamma},\frac{1}{y^{n_\gamma}}\right].
            \label{eq:sigma2r}
        \end{align}
        where $\Gamma$ is the incomplete gamma function.  For $n_\gamma=4$ this is the same result as found in \citep{bib:Bertschinger2006}.  We can furthermore analytically integrate Eq.~\ref{eq:uee} for $x=\eta'/\eta_d$ and $y=\eta/\eta_d$ to obtain:
        \begin{align}
            u_y(x)&=\int_x^y dw \frac{1}{w}\exp\left[-\int_x^w \frac{n_\gamma}{2}\frac{1}{z^{n_\gamma+1}} dz\right] \nonumber \\
            &=\frac{1}{n_\gamma}\exp\left[-\frac{1}{2x^{n_\gamma}}\right] \left[ {\rm Ei}\left(\frac{1}{2x^{n_\gamma}}\right)-{\rm Ei}\left(\frac{1}{2y^{n_\gamma}}\right)\right]
            \label{eq:uee_r}
        \end{align}
        with ${\rm Ei}$ being the exponential integral.  Because only radiation fluctuations source the gravitational potentials at this time, Eq.~\ref{eq:delta_x} is simply the solution and can be written in terms of dimensionless parameters as:
        \begin{widetext}
        \begin{align}
            \delta^r_\chi(y)&=\delta_{\chi}(\eta\rightarrow0)\exp\left[-\frac{1}{2}\epsilon\omega^2 n_\gamma \int_0^y dz \frac{u_y^2(z)}{z^{n_\gamma}}\right]
            +\int_0^y dx \mathcal{S}^r_y(x)\exp\left[-\frac{1}{2}\epsilon\omega^2 \left(\frac{\sigma^2}{\sigma_d^2}u_y^2(x)+n_\gamma \int_x^y dz \frac{u_y^2(z)}{z^{n_\gamma}}\right)\right] \label{eq:delta_xr} \\
            \mathcal{S}^r_y(x)&=3\frac{d\phi^r}{dx}-u_y(x)\left[\frac{n_\gamma}{2}\frac{1}{x^{n_\gamma}}(\eta_d\theta^r_R)+x\omega^2\phi^r\right]
            +\epsilon\omega^2u_y^2(x)\left[\frac{n_\gamma}{2}\left(\frac{\sigma^2}{\sigma_d^2}-x\right)\frac{\delta^r_\gamma+\phi^r}{x^{n_\gamma+1}}-\frac{n_\gamma}{2}\frac{\delta^r_T}{x^{n_\gamma}}-\frac{\sigma^2}{\sigma_d^2}\frac{d\phi^r}{dx}\right]
        \end{align}
        \end{widetext}
        where we use $^r$ to distinguish the radiation era perturbations, $\delta_{\chi}(\eta\rightarrow0)=(-9/10)\Ri$ is the superhorizon matter fluctuation, $\epsilon=T_d/m_\chi\ll1$, and $\omega=k\eta_d$.
        
        The last required ingredient is the source perturbations.  Since WIMP decoupling occurs before neutrinos begin to free stream, anisotropic stress is negligible and so $\phi=\psi$.  We can then combine the Einstein equations (e.g.~\citep[Eqs.~19, 20, 23, 25]{bib:Ma1995}), assuming only radiation perturbations contribute at this time, to obtain \citep{bib:Loeb2005}:
        \begin{align}
            \ddot{\phi}^r+\frac{4}{\eta}\dot{\phi}^r+\frac{k^2}{3}\phi^r=0 \rightarrow \phi^r=3\phi(\eta\rightarrow0)\frac{j_1(\theta)}{\theta}
        \end{align}
        where $\theta=k\eta/\sqrt{3}$, $\phi(\eta\rightarrow0)=(3/5)\Ri$ and we have assumed only an adiabatic growing mode (for large isocurvature modes see \citep{bib:Passaglia2022}, and for the presence of a decaying mode see \citep{bib:Kodwani2019}).  The photon perturbations can then be directly obtained from the Einstein equations:
        \begin{align}
            \delta^r_R=&-2\left[(\theta^2+1)\phi+\theta\frac{d\phi}{d\theta}\right] \\
            \eta_d\theta^r_R=&\frac{3}{2}\theta_d\left[\theta\phi+\theta^2\frac{d\phi}{d\theta}\right]
        \end{align}
        where $\theta_d=k\eta_d/\sqrt{3}$.
        
        \citet{bib:Bertschinger2006} found an extremely accurate approximation for $\delta_\chi$ by solving the moment equations in the limit $\epsilon\rightarrow0$ and then multiplying the resulting density contrast $\delta_{\chi0}$ by a Gaussian damping factor.  This approximation is by construction accurate until $\omega\sim1/\sqrt{\epsilon}=100$, so provided such a scale ends up damped the approximation will always be quite accurate.  Making the same approximation in Eq.~\ref{eq:delta_xr}, $\delta_{\chi0}$ in the radiation era is given by:
        \begin{align}
            &\delta^r_{\chi0}=\delta_\chi(\eta\rightarrow0)\nonumber \\ &+\int_0^ydx\left\{3\frac{d\phi^r}{dx} -u_y(x)\left[\frac{n_\gamma}{2}\frac{1}{x^{n_\gamma}}(\eta_d\theta^r_R)+x\omega^2\phi^r\right]\right\}.
            \label{eq:delta_xr0}
        \end{align}
        When numerically evaluating $\delta_{\chi0}^r$ we assume that for $x\gg1$ the contribution of $\eta_d\theta^r_R$ is negligible, $u_y(x)\simeq\log(y/x)$ and the integral can be performed analytically:
        \begin{align}
            \int_{\theta_a}^{\theta} &\left\{ 3\frac{d\phi^r}{d\theta'}-3\theta'\phi^r\log\frac{\theta}{\theta'}\right\}d\theta'=\nonumber\\ &9\phi_i\left[j_0(\theta')\log\left(\frac{\theta}{\theta'}\right)+{\rm Ci}(\theta')+\frac{j_1(\theta')}{\theta'}-j_0(\theta')\right]_{\theta_a}^{\theta}
        \end{align}
        where ${\rm Ci}$ is the Cosine integral.  We switch to this analytic result at $\eta'=10\eta_d$.  Note that taking $\theta_a\rightarrow0$ and adding $\delta_\chi(\eta\rightarrow0)$ yields the CDM perturbation:
        \begin{align}
            &\delta^r_c=9\phi_i\left[\frac{1}{2}-\gamma_E - \log\theta+{\rm Ci}(\theta) + \frac{j_1(\theta)}{\theta} - j_0(\theta) \right]
            \label{eq:delta_cR}
        \end{align}
        where $\gamma_E$ is the Euler-Mascharoni constant \citep{bib:Bertschinger2006,bib:Voruz2014}.  
        
        Taking $y$ derivatives of Eq.~\ref{eq:delta_xr0}, it is straightforward to convert the integral equation to a closed set of differential equations:
        \begin{align}
            &\frac{d}{dy}\delta^r_{x0}=-(\eta_d\theta^r_{\chi0})+3\frac{d}{dy}\phi^r \\
            &\frac{d}{dy}(\eta_d\theta^r_{\chi0})=-\frac{1}{y}(\eta_d\theta^r_{\chi0})+\omega^2\phi^r+\frac{n_\gamma}{2}\frac{(\eta_d\theta^r_R-\eta_d\theta^r_{\chi0})}{y^{n_\gamma+1}},
        \end{align}
        which is the same as the friction-only fluid equations in \citep{bib:Bertschinger2006} evaluated in the radiation era.  A similar process can be easily applied to the more general Eq.~\ref{eq:delta_x} as well.  An equivalent integral equation for $n_\gamma=2$ (and which can be converted to Eq.~\ref{eq:delta_xr0} via integration by parts) was obtained starting from the moment equations in \citep{bib:Stadler2019}, with $u_\eta(\eta')$ taking the role of the Green's function.  
        
        \begin{figure}
            \includegraphics[width=0.45\textwidth]{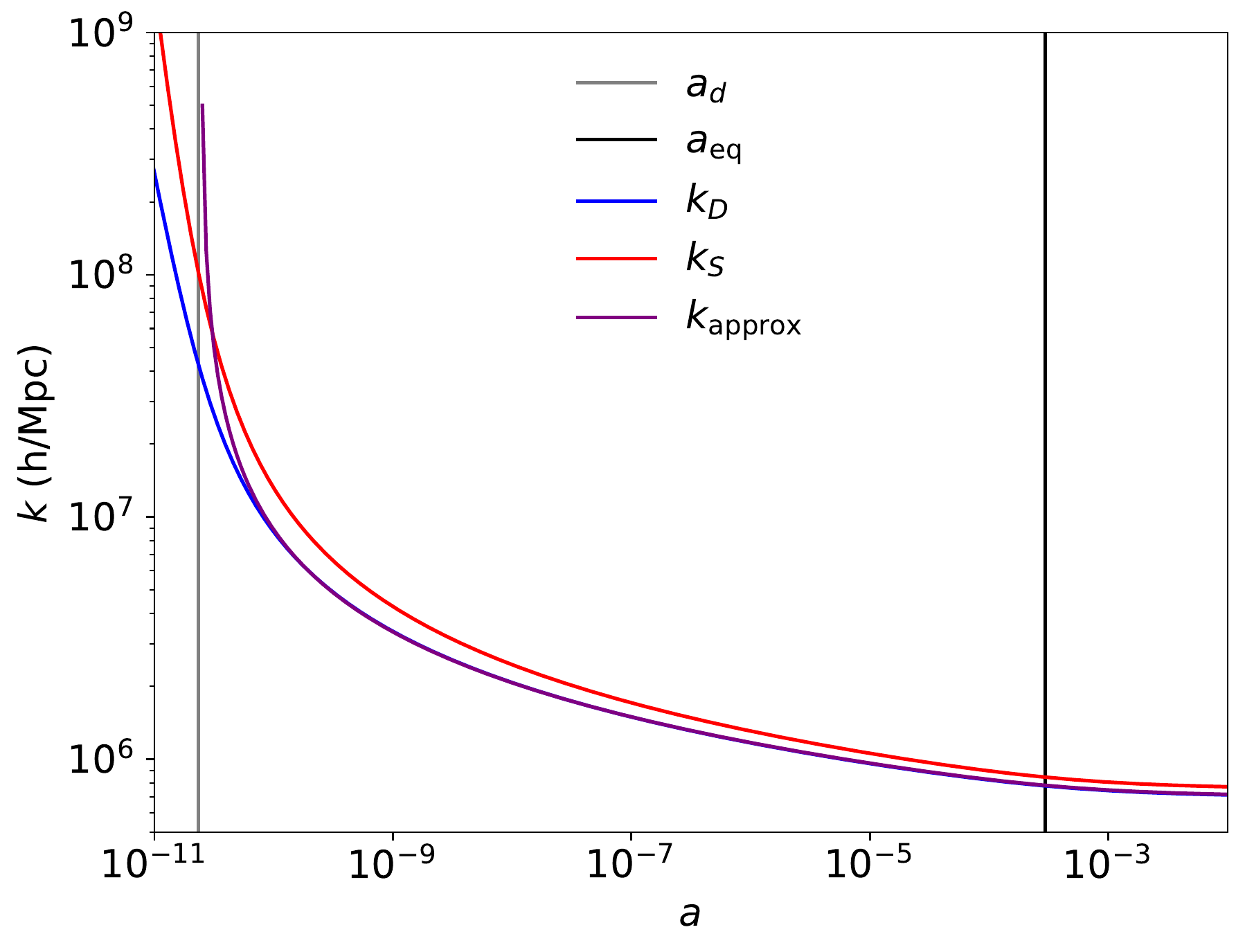}
            \caption{Gaussian damping scales associated with WIMPs decoupling from the cosmic plasma.  The approximate solution matches the integrated diffusion damping scale and rapidly becomes better than 1\% accurate.}
            \label{fig:damping}
        \end{figure} 
        While these equations already include the frictional damping, an additional damping factor is still required.  \citet{bib:Bertschinger2006} found the numerical approximation:
        \begin{align}
            k_{\rm approx}^{-2}=\frac{6}{5}\frac{a_d^2T_d}{m_\chi} \left[ \int_{\eta_\star}^\eta d\eta'/a \right]^2
            \label{eq:kapprox}
        \end{align}
        with $\eta_\star=1.05\eta_d$.
        We show this approximation, alongside Eq.~\ref{eq:kDD} and \ref{eq:kFS} in Fig.~\ref{fig:damping} (note that we extend the calculation to the matter era, as discussed in the next section).  The minimum value of $k_{S}$ occurs around:
        \begin{align}
            x&\simeq\left[\lambda^{-1} W\left(\lambda x^{n_\gamma-1}\right)\right]^{\frac{1}{n_\gamma-1}} \nonumber \\
            \lambda&=2\frac{n_\gamma-1}{n_\gamma}\Gamma\left(\frac{n_\gamma-1}{n_\gamma}\right)
        \end{align}
        with $W$ being the Lambert W function, although in practice we find the precise value numerically.
        We find that $k_{\rm approx}$ matches $k_{D}$, the integrated diffusion damping scale, extremely well after decoupling and therefore consider the following radiation era approximation:
        \begin{align}
            \delta_\chi^{r}(\eta)\simeq\delta^r_{\chi0}(\eta)G_\eta(\eta\rightarrow0).
            \label{eq:delta_xra}
        \end{align}       
        
\begin{figure}
            \includegraphics[width=0.45\textwidth]{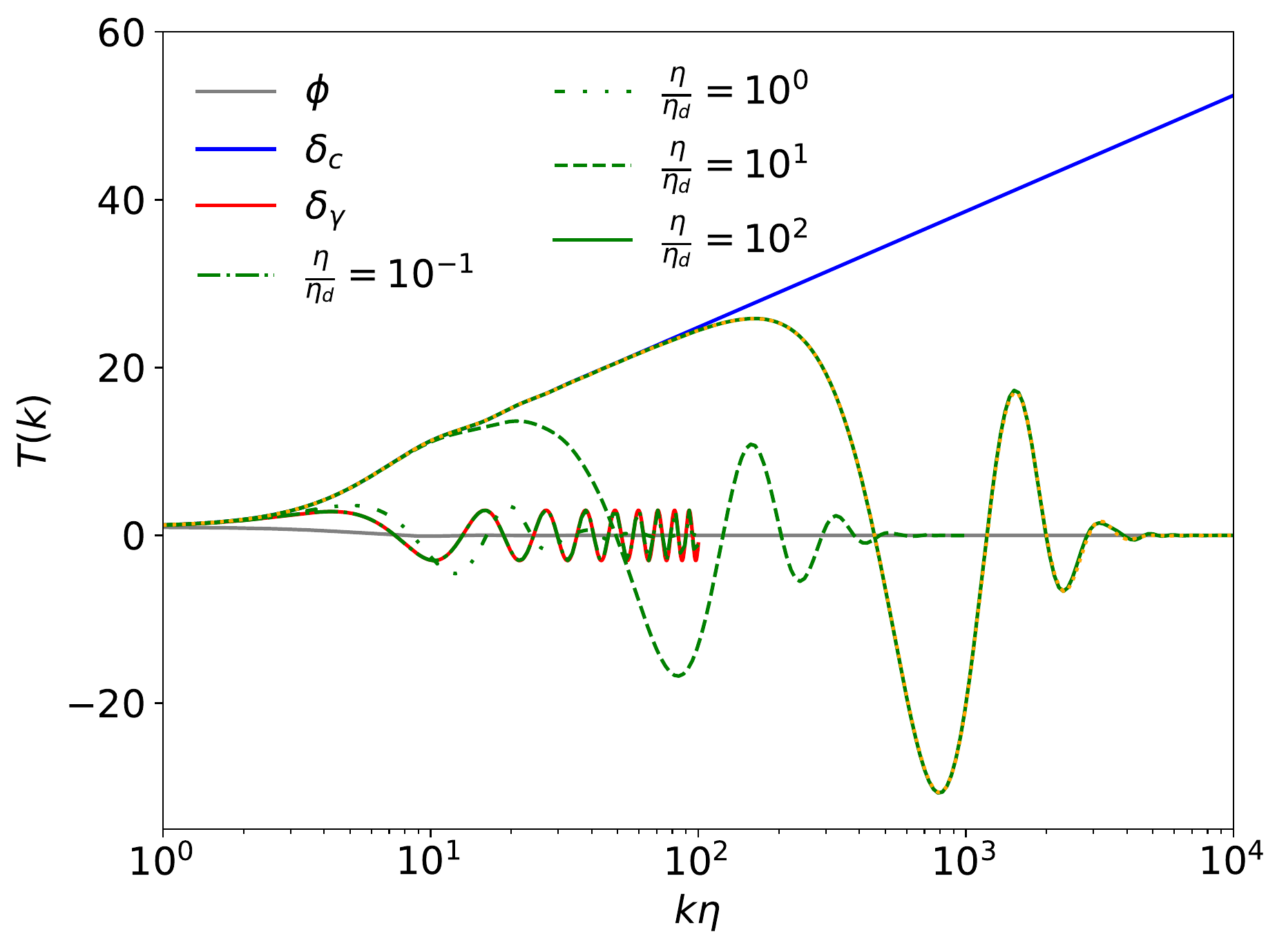}
            \caption{Transfer functions in the radiation era.  Black, blue and red curves show the gravitational potential, CDM density, and photon density.  Green curves show the WIMP density at various times (note that it is separately a function of $\eta/\eta_d$ and $k\eta_d$, not their product).  The dotted orange curve is an approximation to the WIMP transfer function given by Eq.~\ref{eq:delta_xra}.  All transfer functions have been normalized to unity on superhorizon scales.}
            \label{fig:radiation}
        \end{figure}

        We show the resulting transfer functions in Fig.~\ref{fig:radiation}.  WIMP perturbations start tightly coupled to the photons but decouple and then behave like CDM on larger scales, while having acoustic oscillations on smaller ones.  An example of Eq.~\ref{eq:delta_xra} is also shown for $y=10^{2}$.  For our WIMP parameters, the approximation is very accurate.  For example, the third peak only differs from the complete Eq.~\ref{eq:delta_xr} by $\sim1.5$\%, which is comparable to the maximum error quoted in \citep{bib:Bertschinger2006}).  We therefore use it throughout the remaining computations.
    
        \subsubsection{Gilbert's Equation}
        
        We next need to compute the WIMPs evolution after decoupling.  Once decoupled thermally and kinetically, cold dark matter also becomes gravitationally decoupled on subhorizon scales and it evolves only under self-gravity:
        \begin{align}
            -k^2\phi_c=4\pi G a^2\bar{\rho}_c\delta_c
            \label{eq:poisson}
        \end{align}
        even when $\bar{\rho}_R\delta_R\gg\bar{\rho}_c\delta_c$  \citep{bib:Voruz2014}.
        For CDM, this effect leads to the Meszaros Equation \citep{bib:Meszaros1974}; for WIMPs however, the velocity dispersion in Eq.~\ref{eq:sigma2r} asymptotes to a nonzero constant $\sigma^2(y\gg1)\simeq\sigma^2_\infty=\Gamma[(n_\gamma-1)/n_\gamma]\sigma_d^2$ and so free-streaming should in principle be taken into account.  Let us define $\eta_c(k)$ as the time when the WIMP perturbation can be considered subhorizon, fully decoupled and with $\delta_\chi(\eta_c)\simeq\delta_\chi^r(\eta_c)$.  
        Then in addition to the radiation solution (contributing zero on average) there should be an additional contribution from WIMP self-gravity:
        \begin{align}
            &\delta_\chi(\eta)\simeq\delta_\chi^r(\eta)\nonumber\\ &-\int_{\eta_c}^\eta d\eta' u_\eta(\eta') \frac{a}{a_d}\eta_d(k^2\phi_c) \exp\left[-\frac{1}{2}\sigma_\infty^2 \left(\frac{k}{H_r}\right)^2 u_\eta^2(\eta')\right]
            \label{eq:delta_xsplit}
        \end{align}
        where we have set $\psi=\phi$ on subhorizon scales and used the fact that $G_\eta(\eta'\gg\eta_d)$ only has a free-streaming cutoff.
        For $\eta\lesssim\eta_c$, $u_\eta(\eta')$ is given by Eq.~\ref{eq:uee_r} whereas for $\eta\gtrsim\eta_c$:
        \begin{widetext}
        \begin{align}
            u_\eta(\eta') \simeq \frac{1}{n_\gamma}\exp\left[-\frac{1}{2x^{n_\gamma}}\right]\left[{\rm Ei}\left(\frac{1}{2x^{n_\gamma}}\right)-{\rm Ei}\left(\frac{1}{2x_c^{n_\gamma}}\right) + n_\gamma H_r  (\tau-\tau_c)\right],&\ x<x_c \nonumber \\
            \simeq H_r(\tau-\tau'),&\ x>x_c
        \end{align}
        \end{widetext}
        where $x_c=\eta_c/\eta_d$ and we have introduced the superconformal time $a^2d\tau=ad\eta=dt$. 
        After substituting in Eq.~\ref{eq:poisson}, we obtain Gilbert's Equation \citep{bib:Gilbert1966} (see also \citep{bib:Bond1983,bib:Bode2001}):
        \begin{align}
            &\delta_\chi(\tau)=\delta_\chi^r(\tau)+\nonumber\\&\frac{3}{2}f_cH_r^2\int_{\tau_c}^{\tau} d\tau' (\tau-\tau')s\delta_\chi(\tau')\exp\left[-\frac{1}{2}\left(k\sigma_\infty(\tau-\tau')\right)^2\right]
        \end{align}
        where $f_c=\Omega_c/\Omega_m$ and $s=a/a_{\rm eq}$.
        To speed up the calculation we have opted to furthermore take the $\epsilon\rightarrow0$ limit:
        \begin{align}
            \delta_{\chi0}(\tau)=\delta_{\chi0}^r(\tau)+\frac{3}{2}f_cH_r^2\int_{\tau_c}^{\tau} d\tau' (\tau-\tau')s\delta_{\chi0} (\tau')
            \label{eq:delta_xrm0}
        \end{align}
        and then set $\delta_\chi(\eta)\simeq\delta_{\chi0}(\eta)G_\eta(\eta\rightarrow0)$.  We note that this does not appear to be as precise an approximation as in the pure radiation limit.  For instance, the error near the third peak is around $\sim5\%$ at $z=999$.  For our purposes this is acceptable, but in other applications it may not be.  In our calculation we have set $\eta_c={\rm min}\left[\eta,10\eta_d,10\eta_d(k_d/k)\right]$ where $k_d$ is the mode crossing the horizon at decoupling.  We solve Eq.~\ref{eq:delta_xrm0} numerically via trapezoidal integration \citep{bib:Bond1983,bib:Kamionkowski2021}.  
        
        Of course, the validity of these transfer functions rests upon the radiation solution being correct and that WIMPs are the only collapsing matter.  Shortly after WIMP decoupling, at $T\sim1$ MeV, neutrinos also decouple and begin to free stream leading to $\delta_\nu\neq\delta_R$ and, due to neutrino anisotropic stress, $\phi\neq\psi$. Neutrino diffusion also damps photon perturbations on scales $k\gtrsim 5\times 10^4 (T/{\rm MeV})^{2.7}$ \impc{} \citep{bib:jeong2014}, an effect we did not include.
        Immediately following at $T\sim0.5$ MeV is electron-positron annihilation which changes the entropy density, so that the temperature is not inversely proportional to the scalefactor, as well as softens the equation of state.  For CDM perturbations crossing the horizon, \citet{bib:Hu1996} are able to include the effects of neutrino anisotropic stress using semi-analytic expressions.  The effects of changing entropy affect the scalefactor and so could potentially be taken into account just through $u_\eta(\eta'>\eta_c)$.  \citet{bib:Bertschinger2006} was able to approximate the effects of the changing equation of state, finding them to be at the 10\% level.  The last missing effect is other matter: baryons begin to gravitationally collapse after recombination, an effect we do not include.  Further differences can occur if WIMPs are just a single component of a more complex dark sector.  If other matter is also collapsing it would be necessary to retain the free streaming term in Gilbert's equation, as the other matter could source perturbations below the integrated WIMP cutoff.
        
\begin{figure}
            \includegraphics[width=0.45\textwidth]{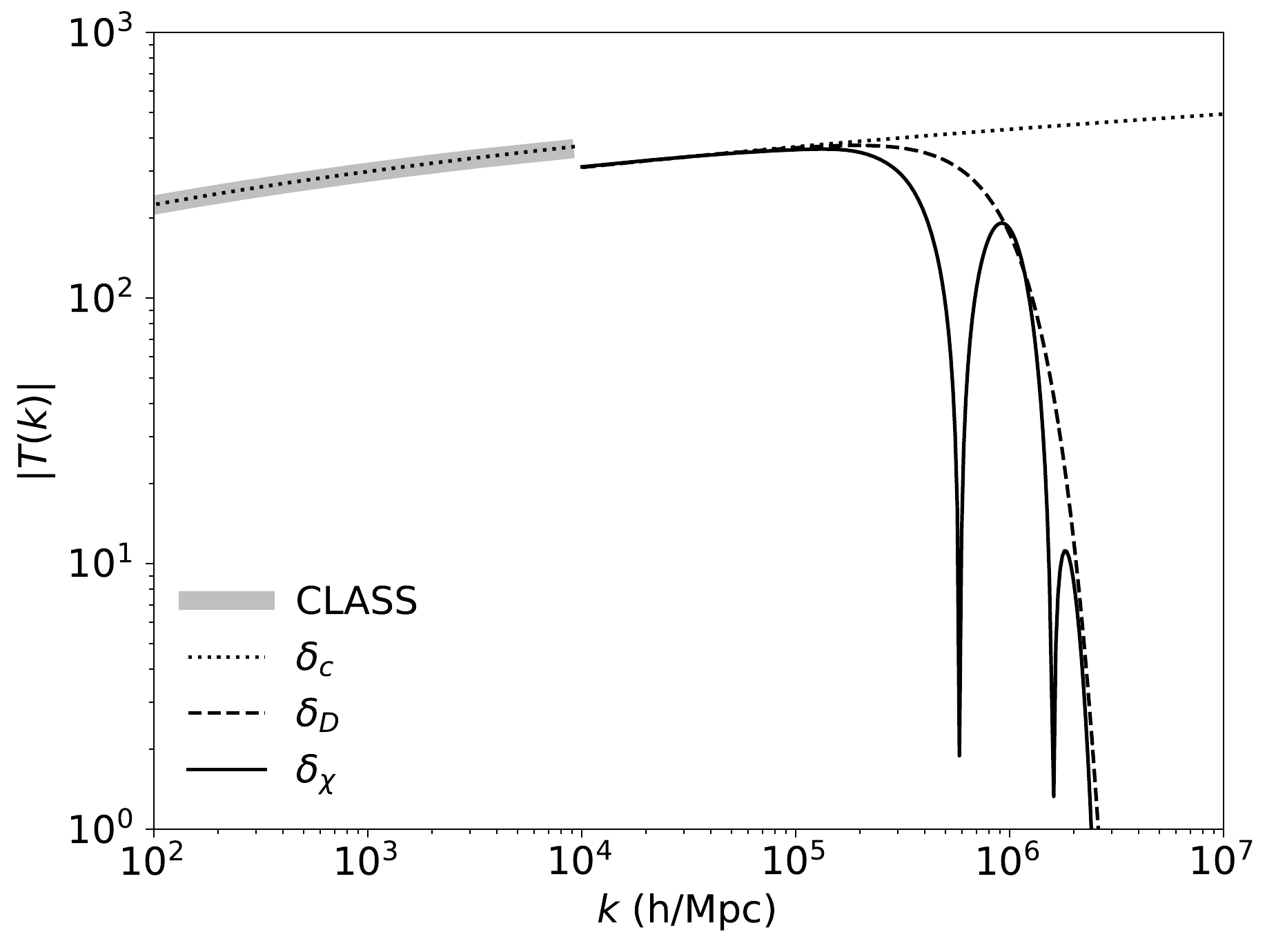}
            \caption{Dark matter transfer functions at $a=10^{-3}$.  The dotted curve is a pure CDM transfer function, whereas the solid line includes the effect of WIMP decoupling.  The dashed curve shows the CDM transfer function multiplied by a Gaussian damping factor.  We also show the CDM transfer function computed with the CLASS Boltzmann code for comparison.  A break in the power spectrum is shown at $\sim10^4$\hmpc{}, a scale characteristic of neutrino decoupling and electron-positron annihilation.  }
            \label{fig:tf}
        \end{figure}

        For our calculation, we have simply neglected these effects to have simpler integration.  In particular, neglecting entropy injection leads to analytic relations for background quantities $a=H_r\eta+(H_r\eta)^2/(4a_{\rm eq})$ and $H_r\tau=\log\left[\eta/(\eta+4a_{eq}/H_r)\right]$ with $H_r=H_0\sqrt{\Omega_r}$.  Both neutrino decoupling and electron-positron annihilation occur at $k\sim10^4$ h/Mpc which is firmly in the range of scales where the WIMP transfer function is the same as CDM.  We therefore use the \citet{bib:Hu1996} approximation with the standard value of the neutrino contribution to the radiation density $f_\nu\simeq0.41$ for scales $k\lesssim 10^4$ h/Mpc, and our calculation for larger wavenumbers.  We show an example transfer function past matter radiation equality at $a=10^{-3}$ in Fig.~\ref{fig:tf}.  On larger scales (where $f_\nu=0.41$) the \citet{bib:Hu1996} approximate transfer function agrees well with those of \class, whereas our calculation based on Gilbert's Equation agrees well with it on on smaller ones (where we set $f_\nu=0$, to take into account the fact that neutrinos were not free-streaming when the mode crossed the horizon).  We thus conclude our calculation yields consistent, albeit imperfect, results for the WIMP density contrast.  Note that the discontinuity in Fig.~\ref{fig:tf} is for illustration only and is never used as an initial conditions for simulations.
        
\subsection{N-body Simulations}
    \label{ssec:nbody}
    After matter-radiation equality, gravitational growth begins and halos can begin to form.  To take such nonlinear evolution into account requires N-body simulations.  We use the ${\rm \sc CUBEP^3M}$ code \citep{bib:HarnoisDeraps2013} which has been modified to evolve ``Particle Dark Matter" starting in the radiation era \citep{bib:Inman2019}.  We use the same high precision parameters to improve gravitational force accuracy: a pairwise force extended over 2 fine cells, a softening length of 1/10 of a fine cell, a logarithmic time step limiter of 0.005, and an offset of up to 16 fine grid cells.  In our simulations, the total number of fine grid cells is $1536^3$ and we employ $2\times768^3$ dark matter particles which are initially placed on a body-centered cubic lattice to reduce discreteness effects \citep{bib:Joyce2005,bib:Marcos2008}.  Initial perturbations are calculated using the Zel'dovich approximation \citep{bib:Zeldovich1970} using both a density and velocity transfer function.  We evaluate the velocity transfer function from the subhorizon continuity equation ($\dot{\delta}+\theta=0$ and note that we take the derivative of $\delta_\chi$, not $\delta_{\chi0}$).  CUBEP$^3$M comes equipped with an on-the-fly spherical overdensity halofinder and we use the halos it identifies based on the virial overdensity ($18\pi^2$, although note that this really only the correct value for the matter era) and with at least $100$ particles.
    
    We consider three types of simulation: CDM, DPS and DAO.  For the pure CDM ($\delta_c$) solution, the cold dark matter transfer function is given by the approximation in \citet{bib:Hu1996}.  The ``damped power spectrum" (DPS, $\delta_D(\eta)=\delta_c(\eta)G_\eta(\eta\rightarrow0)$) simulation uses the CDM transfer function suppressed by the diffusion damping scale.  Note that this only includes the diffusion damping, not the friction damping, and so is an overestimate.  We will use this simulation as a comparison of how different choices of cutoff can impact the results.  Lastly the ``dark acoustic oscillation" (DAO, $\delta_\chi$) simulation which includes both the damping effect and the oscillatory ones associated with decoupling.  On small scales there are two additional effects to consider for WIMPs: the role of thermal velocities, and the effects of artificial fragmentation.  Furthermore, our simulations do not include hydrodynamics and baryons are assumed to be homogeneous throughout the evolution.  We discuss these more in the following subsections, and provide a set of convergence tests in Appendix.~\ref{app:convergence}.
    
    \subsubsection{Thermal Velocities}
        In addition to perfectly cold bulk motions, WIMPs also have thermal velocities leading to dispersive free streaming and scale dependent evolution.  Ideally the simulations would be started well into the matter era when the damping scale has reached an asymptotic constant value; however, because we are considering enhanced primordial power spectra, halos may already be forming at such times.  Thus, we would like to start our simulations as early as possible when perturbations are more linear, but this potentially leads to missed scale dependence.  We start to notice consequential missed scale dependence at around $a=10^{-5}$ and so set this as our initial redshift.  At this redshift the mode crossing the horizon is $\sim0.3$ h/Mpc, and so we can start our larger volume simulations at this redshift as well. 
    
        In addition to the integrated effect of thermal motions in the transfer function, there is also an active suppression of power by thermal motions at any given time due to free streaming.  In the linear evolution, this effect is suppressing power that is already exponentially damped, and so is not as important as the integrated effect.  However, nonlinear evolution transfers power from large scales to small scales due to mode-coupling \citep{bib:Crocce2006}, a process which thermal velocities could inhibit.  Ideally, one would solve the collisionless Boltzmann equation directly, but such simulations have only recently become possible on the largest supercomputers due to the $\mathcal{O}(N^6)$ scaling \citep{bib:Yoshikawa2021}.  A common approach to take thermal velocities into account using standard N-body methods, used for both warm dark matter simulations (e.g.~\citep{bib:Bode2001,bib:Colin2008,bib:Maccio2013,bib:Leo2017}) and simulations including hot dark matter in the form of neutrinos (e.g.~\citep{bib:Brandbyge2008,bib:Viel2010,bib:Inman2015}), is to add a random velocity drawn from $f_0(v)$ (or some compensated distribution, e.g.~\citep{bib:Bird2018,bib:Elbers2021}) to each particle.  We find that adding random velocities does not work well here as they induce random correlations which immediately lead to completely unacceptable fragmentation.  
    
        To avoid such Poisson noise, one can introduce regularity in velocity space as was done for neutrinos in \citet{bib:Banerjee2018}.  We have tested a much simplified version of this method, using just a single shell of velocity, and find that random structures do not form.  While a single shell does not capture the full impacts of thermal motions (as some particles will be much hotter, and some much colder), it does allow us to qualitatively test whether our results are affected by free streaming.  More details, alongside convergence tests with respect to initial redshift and thermal velocities, are given in Appendix~\ref{app:con:tv}.
    
    \subsubsection{Artificial Fragmentation}
        The other numerical effect associated with a cutoff is known as artificial fragmentation.  Below the cutoff scale there are no physical perturbations, but there are numerical ones which begin to grow and fragment in filaments \citep{bib:Wang2007}.  It is observed in hot dark matter simulations \citep{bib:Wang2007}, warm dark matter simulations \citep{bib:Bode2001,bib:Lovell2012}, cold dark matter simulations without enhanced power \citep{bib:Ishiyama2020}, ultracompact minihalo simulations \citep{bib:Delos2018a}, as well as simulations of DAOs in ETHOS simulations \citep{bib:Lovell2018}. Thus, we expect our simulations to suffer from this fragmentation even if the specific shape of the power spectrum may be different from those cases.  
        
        The principle effect is small halos forming along filaments and halo mass functions that do not have the expected cutoff below the mass scale associated with the cutoff.  For hot dark matter, the mass scale associated with this fragmentation is $M_{\rm lim}\simeq10.1\bar{\rho}dk_p^{-2}$ with $d=L/N_p^{1/3}$ being the interparticle spacing, $k_p$ being the peak of the power spectrum and $\bar{\rho}$ being the mean density \citep{bib:Wang2007} and this formula gives a reasonable approximation in ETHOS based DAO simulations as well \citep{bib:Lovell2018}.  In our work there is some ambiguity as to where the peak of the power spectrum is as it depends on redshift and whether acoustic oscillations are included, but our tests indicate $M_{\rm lim}$ is consistent here as well.

        Given that such artificial halos affect the halo mass function, the next question is how to avoid them.  One option is to filter them out, based on criteria such as convergence in Lagrangian space \citep{bib:Lovell2014} or virialization \citep{bib:Agarwal2015}.  Alternatively, one can attempt to stop them from forming by reducing the force resolution of the simulation to match the mass resolution \citep{bib:Melott1989}, as it has been shown that lower resolution simulations like pure particle-mesh reduce the fragmentation \citep{bib:Angulo2013}.  A more advanced numerical method which interprets particles as tracers of the continuous CDM phase sheet could also be employed \citep{bib:Stucker2020,bib:Stucker2022}.  We explore the effect of force and mass resolution in Appendix \ref{app:con:af}.  However, reducing force resolution to deal with smaller halos also reduces our ability to study heavier high mass halos as well.  We therefore defer a detailed study of the lower end of the halo mass function to a future study, and instead focus on halos that are well resolved by the simulation.
        
    \subsubsection{Baryonic Effects}
        Before recombination baryons are coupled to the CMB and so their perturbations may be safely neglected on scales relevant for our simulations.  After recombination however, they begin to gravitationally collapse into dark matter halos.  Unlike WIMPs, baryons remain collisional and have some pressure support to prevent collapse on such small scales \citep{bib:Naoz2007}.  It is therefore not unreasonable to treat them as homogeneous on very small scales and at very early times.  However, at later times when bigger halos are forming it becomes a much worse approximation as baryons do collapse and begin to form stars and (proto-)galaxies.  An accurate treatment of this would require hydrodynamical simulations including high redshift chemistry \citep{bib:Hirano2015}.  
        
        In both the matter and radiation eras, a component being homogeneous leads to a reduced growth factor \citep{bib:Bond1980,bib:Inman2019}.  Thus, when baryons catch up to CDM, which may be different with enhanced structure, will affect structure formation.  We therefore have performed a simple test where we assume that instead of being homogeneous, baryon perturbations are exactly the same as the WIMP ones starting at some time after recombination.  Note however that this doesn't take into account the fact that on some scales baryons become more clustered than CDM as they have the ability to cool \citep{bib:Chisari2018}.  The results of these tests are given in Appendix~\ref{app:con:bc}, and demonstrate that our results are an underestimate of the true clustering.
        
\section{Results}
    \label{sec:results}
\begin{figure}
        \includegraphics[width=0.45\textwidth]{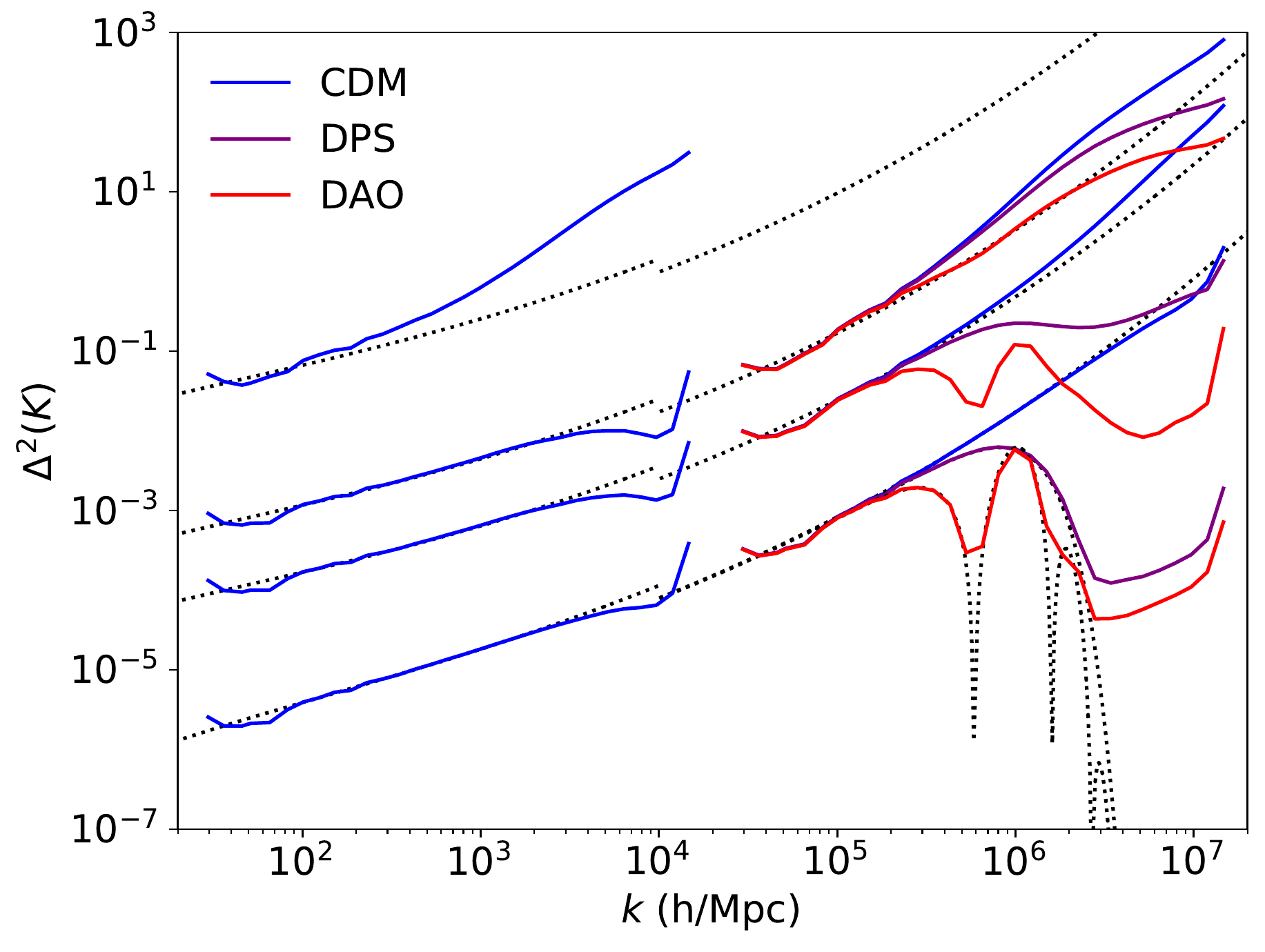}
        \caption{Power spectra for WIMP dark matter with enhanced power on small scales.  Dotted curves show linear transfer functions at the initial conditions $z_i=99999$ (CDM, DPS, DAO) and after evolution at $z=999,299,29$ (just CDM).  The blue curves shows the pure cold dark matter model, the purple curves show a model with an initial Gaussian cutoff in the power spectrum, whereas the red curves are the case including full decoupling.  Note that the small volume (large wavenumber) simulation is run only to $z=299$.}
        \label{fig:power}
    \end{figure}

    With the tools developed in the previous section, are now able to examine the formation of WIMP halos with enhanced small-scale power.  We have run two classes of simulations to focus that focus on different scales and redshifts.  The first are in $(300{\rm\ h^{-1}pc})^3$ volumes, evolve from $a=10^{-5}$ until $z=299$ and are focused on the formation of very high redshift halos near the cutoff in the power spectrum.  The second set of simulations is run in $(300{\rm\ h^{-1}kpc})^3$ volumes until $z=29$ in order to understand how the increase in power may affect the formation of the halos that will host the first stars and galaxies.  We show power spectra from our set of simulations in Fig.~\ref{fig:power}.  On all scales, nonlinear evolution begins substantially earlier than is typically assumed.
    
    \subsection{Halos at $z\sim300$}
            
\begin{figure*}
            \includegraphics[width=1\textwidth]{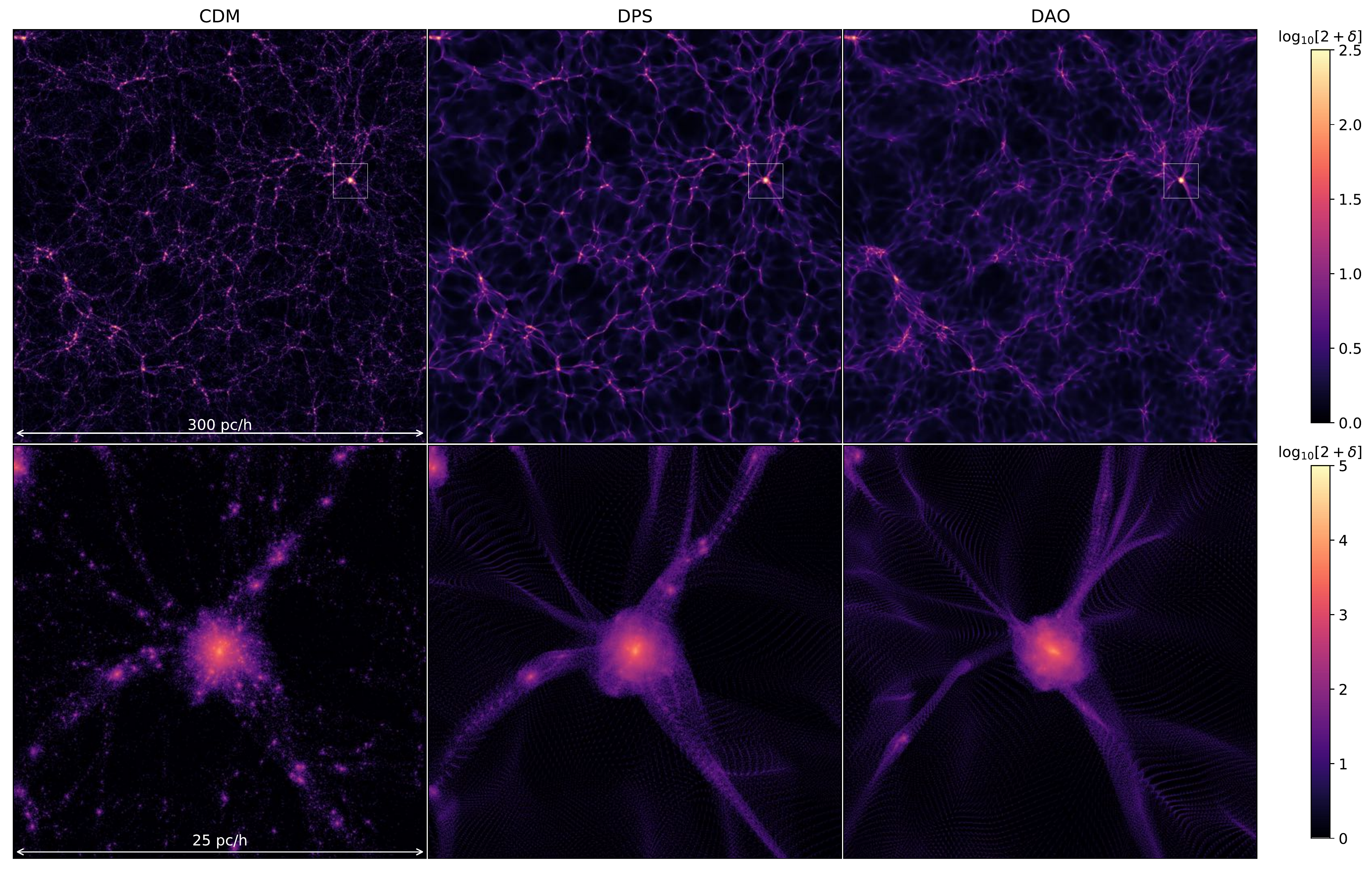}
            \caption{Slices of dark matter density at $z=299$ for CDM, WDM and DAO initial perturbations.  The white box in the top panels is zoomed in on with $8\times$ resolution in the bottom panels and shows the largest halo in the simulation.  Without the enhanced small-scale power, the density field would still be linear at this redshift.}
            \label{fig:6slice}
        \end{figure*}

        We now consider the very early halos that form in our small volume simulation.  The top row of Fig.~\ref{fig:6slice} shows density slices from the small volume simulations at $z=299$.  We see that a substantial amount of structure has already formed in all three models.  However, by eye we can see that the CDM simulation has substantially more structure than the other two.  Furthermore, the DPS simulation is clearly more clustered than the DAO one.  This can also be seen very easily in Fig.~\ref{fig:power}: while the DPS simulation is catching up to the CDM one, the DAO one has yet to do so.  The difference we see does have a simple interpretation: there is substantially more power in the DPS simulation even in linear theory.  For instance, the variance, $\int \Delta^2 d\log k$ is $1.76\times$ larger in the DPS simulation than the DAO one.  This is not the case with no running parameters, where it differs by $1.13\times$ without running.  The precise shape of the power spectrum near the peak has a substantial effect and approximate damping scales may not lead to accurate conclusions.  While we might expect that these differences will further diminish at later redshifts, a substantial delay may be sufficient to reduce the constraining power of the CMB.  We furthermore find the oscillatory features present in the initial power spectrum are removed by nonlinear evolution, similar to the results found for ETHOS models with DAO at larger wavenumbers \citep{bib:Schaeffer2021}.  Lastly, we note that we ran a DAO simulation without running and found that the power spectra remains linear and the halofinder does not find any halos at this redshift.
        
\begin{figure}
            \includegraphics[width=0.45\textwidth]{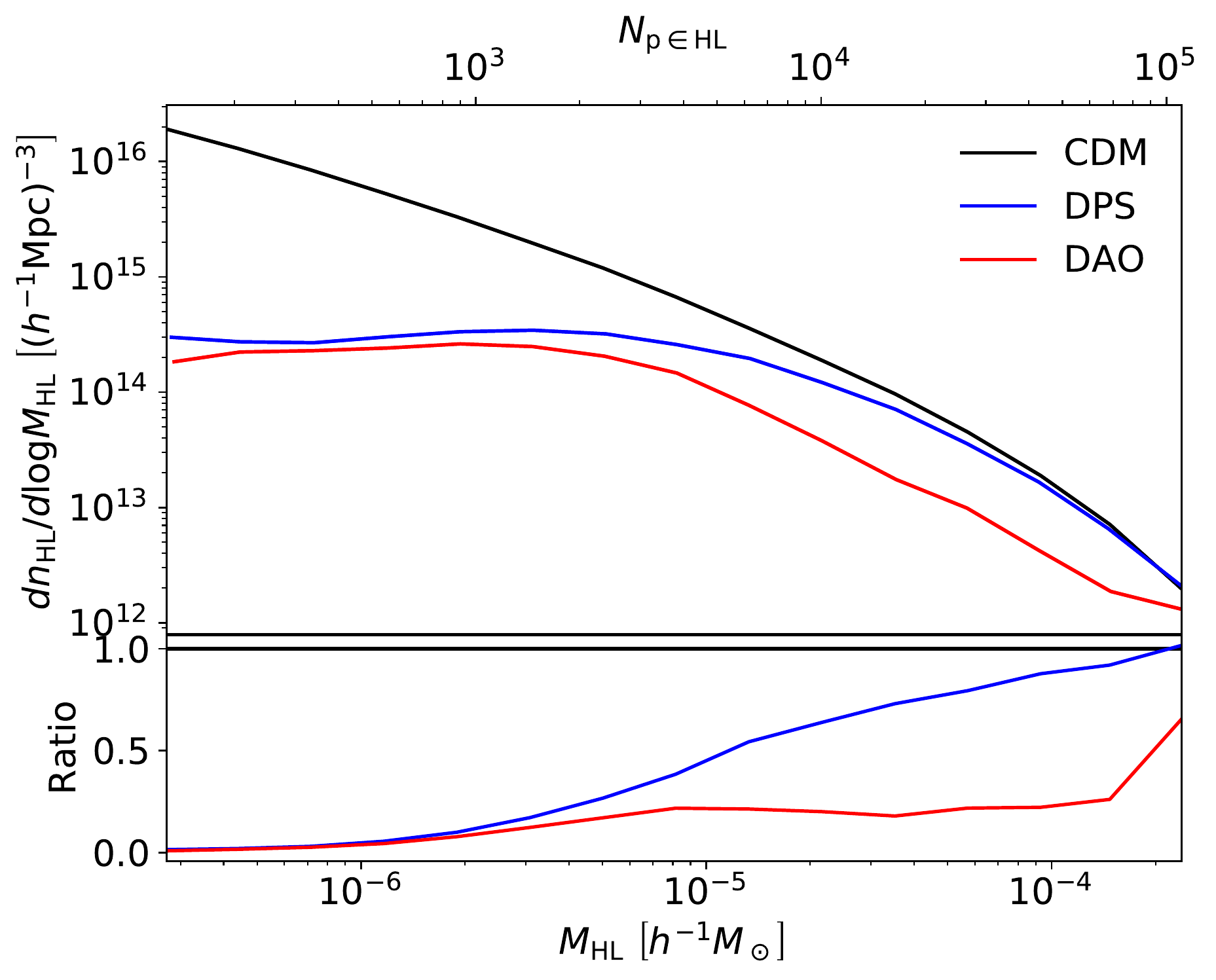}
            \caption{Halo mass function at $z=299$.  A cutoff in the power spectrum for both the DPS and DAO simulations leads to a cutoff in the halo mass function.  The bottom panel shows the ratio with respect to CDM in each bin.}
            \label{fig:hmf}
        \end{figure}

        We can further quantify the differences between the three simulations by considering the halo mass function, which we show in Fig.~\ref{fig:hmf}.  Comparing the DPS and CDM simulation, we see a characteristic suppression at $\sim10^{-5} {\rm\ h^{-1}M_\odot}$.  The DAO simulation also is suppressed, but on all scales, which is consistent with the lower power spectrum.  While there isn't a visible uptick in the damped halo mass functions due to artificial halos, our convergence tests (see Fig.~\ref{fig:hmf_u}) suggest that this is due to limited resolution.  
        The bottom panel of Fig.~\ref{fig:hmf} shows the ratio of the DPS and DAO mass functions to the CDM one (note that this is done with the same mass bins, but plotted at the mean CDM mass per bin).  We note that ETHOS models with DAO have oscillations in the halo mass function \citep{bib:Schaeffer2021}, which could be possible here, although it is difficult to tell with our resolution.

\begin{figure}
            \includegraphics[width=0.45\textwidth]{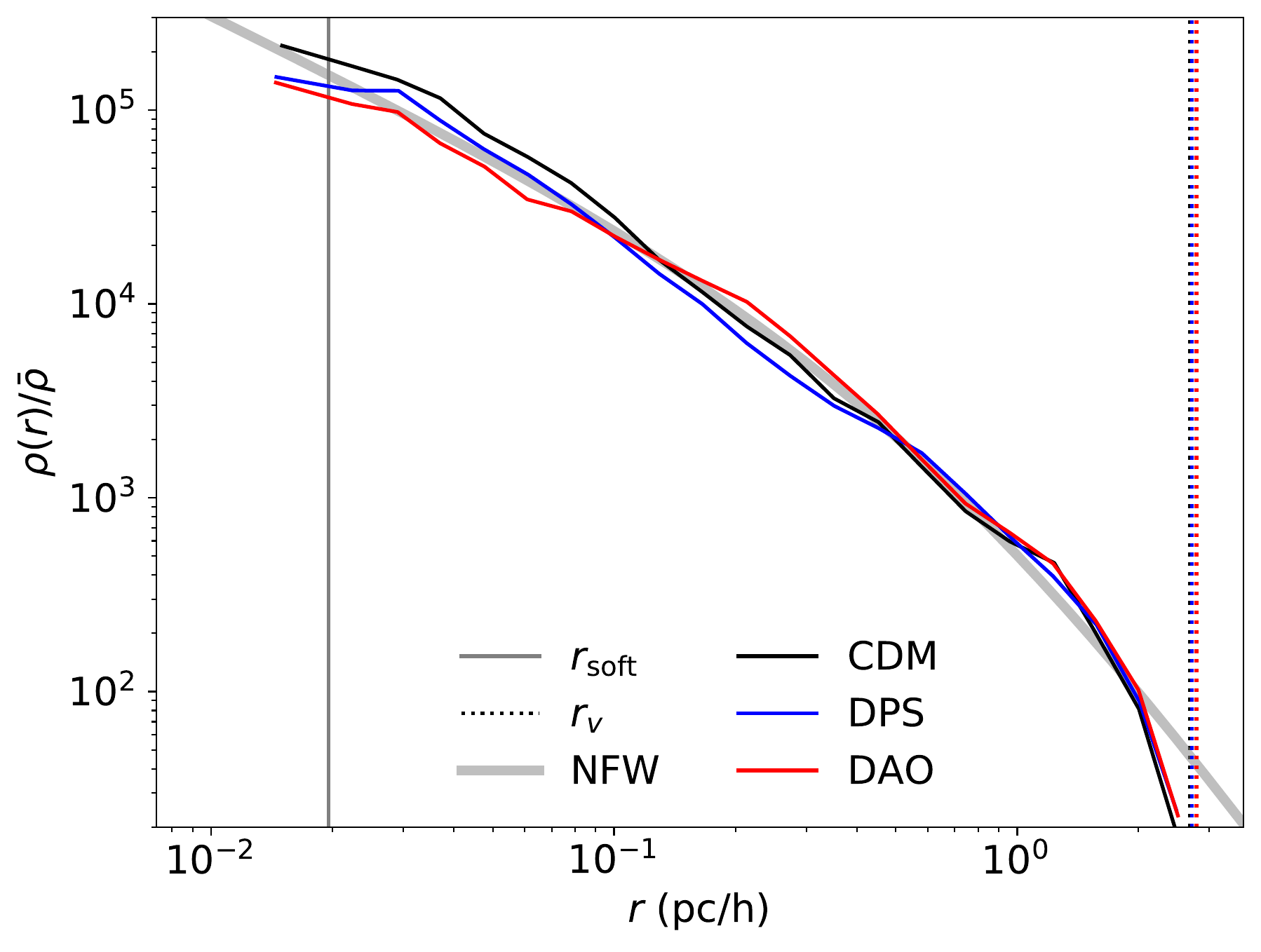}
            \caption{Density profile of the largest halo at $z=299$.  Regardless of initial conditions the profile remains the same on all resolved scales is matched well by an NFW density profile.  Vertical lines indicate the force softening length and the halo virial radii.}
            \label{fig:rho}
        \end{figure}

        We lastly consider the interior of halos.  In the bottom row of Fig.~\ref{fig:6slice} we show a zoomed in region around the largest halo in our simulation, with mass $10^{-3} {\rm h}^{-1}{\rm M}_\odot$ resolved by $\sim5\times10^5$ particles.  Surprisingly, the halo is heaver in the DAO simulation and lighter in the CDM one.  We show the density profile of particles within the virial radii in Fig.~\ref{fig:rho}, and find it is very similar in all three simulations.  It furthermore agrees well with a Navarro-Frenk-White (NFW) profile with concentration $c\sim7.5$ \citep{bib:Navarro1997}.  Let us now assess whether this halo is consistent with those found in simulations of ultracompact minihalos.  \citet{bib:Delos2018a} found that halos forming from extremely rare peaks in the density field have interior slope $\rho\propto r^{-3/2}$ instead of NFW.  However, because it formed out of such a large peak their halo collapsed at $a\sim10^{-3}$ as the very first halo.  We inspected earlier checkpoints of our simulation and find that our halo forms by mergers of smaller halos at $z\sim500$.  It therefore makes sense that it has the relaxed NFW profile instead.  This is also consistent with the boosted simulations of \citet{bib:Gosenca2017}, where NFW profiles are also found.  They reported substantially higher concentrations ($c\gtrsim100$) at lower redshifts, which could also be the fate of the halos in our simulations given typical concentration evolution \citep{bib:Bullock2001}.  Thus, our results appear compatible with previous numerical simulations of peaked primordial power, even though we consider much smaller scales.  Furthermore, this picture is also consistent with the first halos in standard $\Lambda$CDM cosmology without running \citep{bib:Ishiyama2014}.  Due to resolution we have focused only on a heavy halo; however, the first halos may have steeper profiles than NFW \citep{bib:Ishiyama2014,bib:Delos2018a,bib:Delos2022} making them an important future target.
        
    \subsection{Halos at $z\sim30$}

        \begin{figure*}
            \includegraphics[width=1\textwidth]{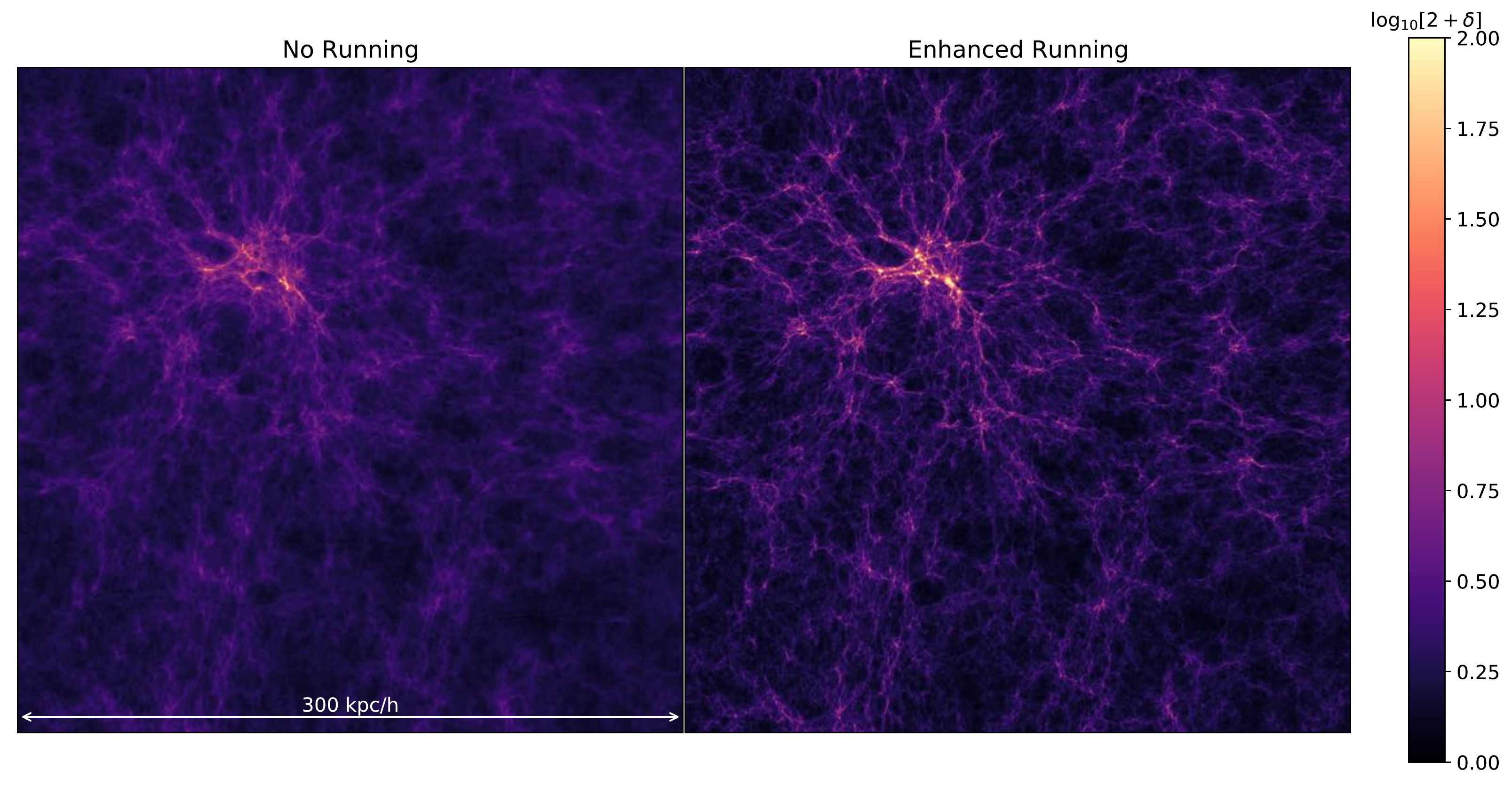}
            \caption{Slices of dark matter density at $z=29$ for initial power spectra with no running and enhanced running parameters.  Structures are substantially more developed with running due to the increased small-scale power.}
            \label{fig:2slice}
        \end{figure*} 

        At later times, much larger halos can begin to form, and we can study them with our larger volume simulations.  In order to prevent being biased by a single realization, we ran five simulations with different random seeds and tabulate the number of heavy halos at $z=29$ in Table~\ref{tab:halomass}.  The first simulation listed has the same random seed as the simulations used for the smaller volumes, and it does not seem particularly unusual.  We find in general that there are hundreds of halos with $M\geq10^4$\hmpc{} and a small number with $M\geq10^5$\hmpc{}.  The largest halo we found has a mass of $3.2\times10^5$\hmpc, which we show in Fig.~\ref{fig:2slice}.
    
        In order to understand the effect of the enhanced running, we ran a couple simulations with the same random seeds as before, but with no running in the primordial power spectra.  We find that there are essentially no heavy halos at this time in any simulation.  This can clearly be seen by comparing the left and right panels of Fig.~\ref{fig:2slice}.
    
        \begin{center}
        \begin{table}
        \begin{tabular}{r | c | c | c }
            \multicolumn{1}{c}{}&\multicolumn{2}{|c}{Halos with $M\ge$}&\multicolumn{1}{|c}{Max Mass} \\
            Simulation \# & \multicolumn{1}{|c}{$10^4 {\rm h}^{-1}{\rm M}_\odot$}&\multicolumn{1}{|c|}{$10^5 {\rm h}^{-1}{\rm M}_\odot$}&$(10^5{\rm h}^{-1}{\rm M}_\odot$) \\
            \hline
            Enhanced Running 1 & 245 & 2 & 1.3 \\
            2 & 303 & 0 & 0.8 \\
            3 & 436 & 8 & 3.2 \\
            4 & 339 & 1 & 1.1 \\
            5 & 326 & 1 & 1.8 \\
            No Running 1 & 0 & 0 & 0.06 \\
            3 & 8 & 0 & 0.4
        \end{tabular}        
        \caption{Number counts of large halos at $z=29$ in volumes of $(300{\rm\ h^{-1}kpc})^3$.  With enhanced running, halos large enough to have stars form when they otherwise would not.  Simulations with the same \# have the same initial seeds.}
        \label{tab:halomass}
        \end{table}
        \end{center}

\section{Discussion}
    \label{sec:discussion}
    Having established that halo formation can occur much earlier than is typically assumed, we now discuss potential consequences of the enhanced power spectrum.  One chief difference between WIMP dark matter and pure cold dark matter is the WIMPs ability to annihilate.  If the resulting particles are gamma rays then strong constraints can be placed on WIMPs based on observations of annihilation in the late Universe \citep{bib:Nakama2018}.  Alternatively, energy injection into the baryonic gas at much earlier times can be used to constrain WIMPs using the CMB \citep{bib:Kawasaki2022} or global 21 cm measurements \citep{bib:hiroshima2021}.  If, instead, one assumes the dark matter is WIMPs, then constraints on the primordial power spectrum can be placed \citep{bib:Gosenca2017,bib:Delos2018b}.
    
\begin{figure}
        \includegraphics[width=0.45\textwidth]{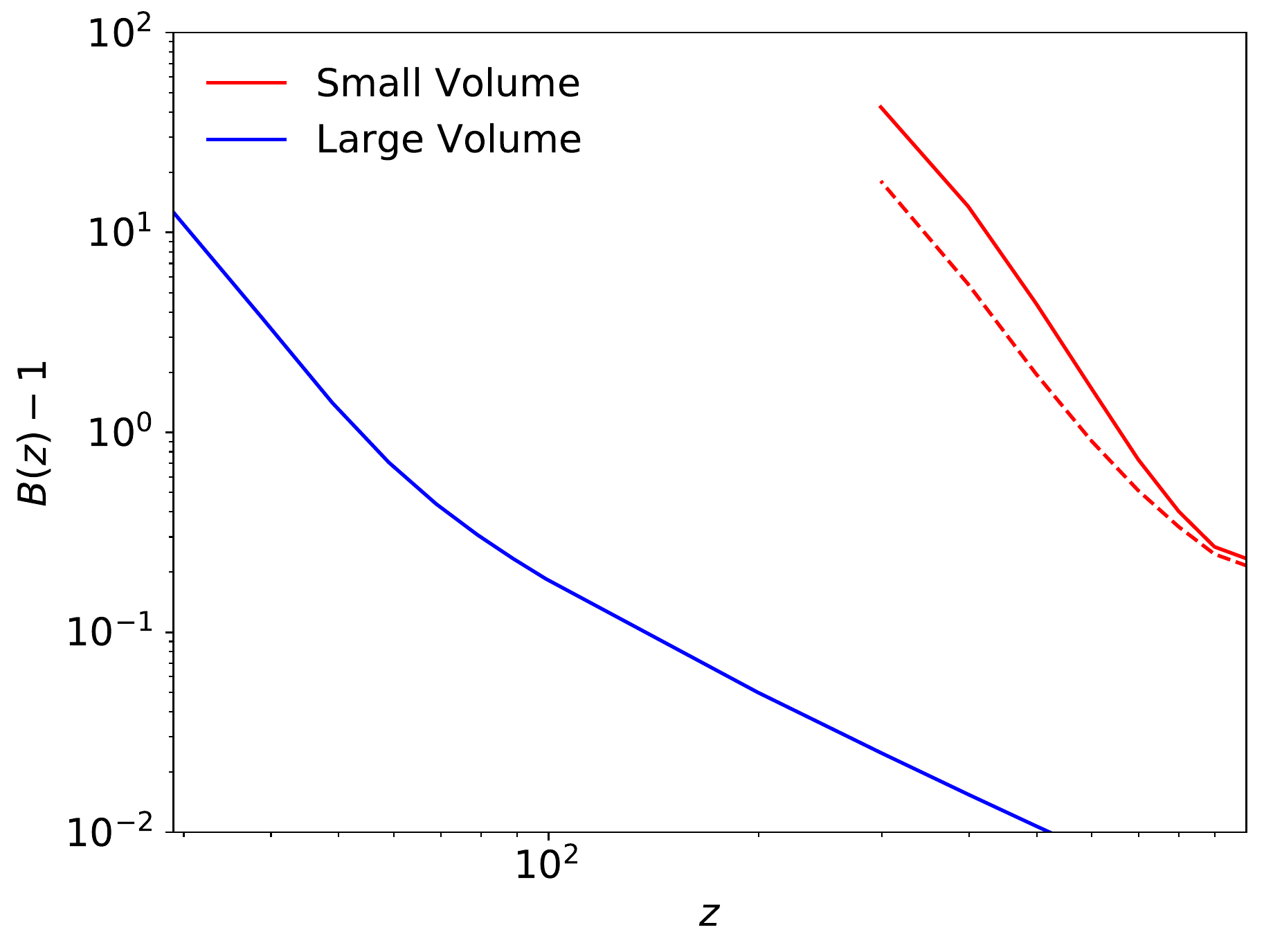}
        \caption{Cosmological boost factor computed over the range of scales and redshifts covered by the two volumes of our simulations.  The dashed curve computes the boost factor only until $k=5\times10^6{\rm\ h^{-1}Mpc}$ instead.  Without running, the boost factor is $\sim10$ at $z\sim30$ \citep{bib:Takahashi2021}.}
        \label{fig:boostSL}
    \end{figure}

    Since annihilation is proportional to the squared density, a simple way to quantify the effect is through the cosmological boost factor, which can be computed as an integral over the power spectrum \citep{bib:Serpico2012,bib:Takahashi2021}:
    \begin{align}
        B(z)=1+\int \Delta^2_\chi(k) d\log k.
    \end{align}
    In principle the integral is eventually cutoff by some physical process such as the annihilation of the interior of a halo \citep{bib:Bringmann2012}.  Because we only have a finite dynamic range, we instead consider only the scales covered by the two volumes simulated.  We show these two boost factors integrated to the particle Nyquist frequency ($\pi N_p^{1/3}/L$) in Fig.~\ref{fig:boostSL}.  For the small volume simulation, we also the result with integration to $k=5\times10^{6}{\rm\ h^{-1}Mpc}$ as a dashed curve, which should partially remove the small-scale noise seen in Fig.~\ref{fig:power}.  For comparison, \citet{bib:Takahashi2021} ran simulations covering nonlinear evolution on scales $10^1\lesssim k/({\rm h/Mpc})\lesssim10^8$ without running and their boost factor was only $B\sim10$ at $z\sim30$.  We therefore conclude that with enhanced small-scale power there will be a substantial enhancement in WIMP annihilation after recombination and continuing to the formation of first galaxies.

    Of course, because both the annihilation rate and the primordial power spectra are proportional to $\rho^2$, there is a degeneracy that prevents concrete constraints on either WIMPs or $\Delta^2_\Ri$.  It is interesting therefore to look for other ways of constraining an enhanced power spectrum, which generally leads to looking for the impact of this power spectrum on baryonic structures instead of thermal properties.  While there is potential to constrain small-scale baryonic perturbations at high redshifts via the CMB \citep{bib:Lee2021}, a more direct probe is how the first stars and galaxies are formed.  As we have shown in Table~\ref{tab:halomass}, the enhanced power spectrum leading to a peak at $k\sim10^6 {\rm\ h^{-1}Mpc}$ also increases the number of larger mass halos at later times.  While our simulations don't have the necessary hydrodynamics to study this explicitly, the largest halos in our large volume simulations have masses comparable to the star forming halos in \citet{bib:Hirano2015} (see their Table~1) which were run with a value of $\sigma_8$ increased by $1.5$.  We therefore conclude that it is very plausible that the formation of first stars and galaxies will be affected.  Of course, similar to the very first halos, we expect the very first stars to form in very rare peaks of the density field.  One could study such rare halos by finding an initial random field with a large over density analogously to \citep{bib:Gosenca2017,bib:Delos2018a}.
    
    There are also a number of other uncertain processes that could occur at these redshifts, such as the formation of super-massive black holes \citep{bib:Woods2019} and the potential origin of magnetic fields through structure formation \citep{bib:Naoz2013}, which may be changed by the increased structure formation of a blue-tilted power spectrum.  If these early forming halos can survive until later times as subhalos \citep{bib:Delos2019}, additional types of constraints are possible due to their gravitational influence.  For instance, higher density halos produced by an enhanced power spectrum can lead to potentially detectable signals from astrometric weak gravitational lensing \citep{bib:VanTilburg2018} or through their impact on dark matter substructure \citep{bib:Ando2022}.

\section{Conclusion}
    \label{sec:conclusion}

    We have considered a viable cosmological scenario in which the first Earth mass halos form much earlier than they do when large-scale $\Lambda$CDM is extrapolated to small scales.  In the linear regime we solved the Boltzmann-Fokker-Planck equation to obtain a realistic estimate of how the matter power spectrum is cut-off on small scales by WIMP decoupling from the cosmic plasma.  We then used this solution as initial conditions for N-body simulations to study halos in the nonlinear regime.  We have found that early nonlinear evolution can lead to substantially increased annihilation signatures at early redshifts.  We also found that much larger and potentially star forming halos can form at earlier times as well if the enhancement to the primordial power spectrum occurs over a broad range of scales.  The next goal is to turn these qualitative conclusions into specific constraints on the primordial power spectrum and dark matter microphysics. 
    
    However, there are many important physical processes that we have neglected in our calculation.  In the transfer function, we did not include the effects of neutrino decoupling and electron-positron annihilation.  In our N-body simulations, we do not include relic thermal velocities nor the growth of baryonic perturbations.  We also find evidence of artificial halos at similar mass scales to the predicted first halos, making their study challenging with our simulation resolution.  These deficiencies do not appear impossible to solve, and improvements in each case would certainly be worthwhile.
    
    Lastly, we have only considered a single set of WIMP parameters $\{m_\chi,T_d,n_\gamma\}$ and running parameters $\{\alpha_s,\beta_s\}$.  It would be both interesting and useful to study how varying these parameters may affect the early Universe.  For instance, larger perturbations (either from a further enhanced power spectrum or by faster decoupling, $n_\gamma\gg1$ \citep{bib:Kamada2018}) could lead to more energy injection closer to recombination, from which CMB constraints can be placed \citep{bib:Kawasaki2022}.  Alternatively, changing the WIMP mass or decoupling temperature can lead to a different minimum halo mass and formation time.
    
\section{Acknowledgements}
    We acknowledge valuable discussions with Tobias Binder and Naoki Yoshida.  This work was supported in part by JSPS KAKENHI Grant Numbers JP17H01131 (K.K.), and 
    MEXT KAKENHI Grant Numbers JP19H05114, JP20H04750, and JP22H05270 (K.K.).  Kavli IPMU is supported by World Premier International Research Center Initiative (WPI), MEXT, Japan.  This research made use of {\sc NumPy} \citep{bib:NumPy2020}, {\sc SciPy} \citep{bib:SciPy2020}, {\sc Matplotlib} \citep{bib:Matplotlib2007} and NASA's Astrophysics Data System Bibliographic Services.

\bibliographystyle{apsrev}
\bibliography{thebib}

\appendix

\section{Simulation Convergence}
\label{app:convergence}
    In this section we present convergence tests of the results presented in Section~\ref{sec:results} of various numerical issues discussed in Section~\ref{ssec:nbody}.

    \subsection{Thermal Velocities}
    \label{app:con:tv}
        To test the impact of thermal velocities on the small volume simulation, we have implemented a much simplified version of the method presented in \citet{bib:Banerjee2018}.  Instead of using many shells to sample $f_0(v)$, we instead use just a single representative shell.  We replace each particle by a set of 6 particles, each given the same velocity $u$ but oriented in the six directions (e.g.~$+x,-x,+y,-y,+z,-z$) of the initial lattice.  The velocity, $u$, should be representative of a shell $f_0(v)=(1/4\pi u^2)\delta_D(v-u)$.  Such a shell distribution behaves similarly to the Gaussian distribution except with a damping term $j_0(ku(\tau-\tau'))$ instead of $\exp\left[-(k\sigma_\infty(\tau-\tau'))^2/2\right]$ \citep{bib:Inman2017,bib:Bird2018}.  Of course, there is no perfect value of $u$ to match the two functions, but we can match the first two coefficients of the Taylor expansions by using $u=\sqrt{3}\sigma_\infty$ and a slightly higher than average value is conservative for our convergence test.  In principle, individual shells should have distinct transfer functions \citep{bib:Inman2020}, but for this test we simply use the DAO density and velocity transfer functions.  Furthermore, we are neglecting any perturbations in the velocity dispersion, which is expected to introduce errors $\sim\delta$ \citep{bib:Brandbyge2017}.  As a test of the method, we have also considered the case where we use $u=\sqrt{300}\sigma_\infty$, which could model WIMPs with $\epsilon=10^{-2}$ (although with a different transfer function).
    
        To prevent artificial forces of particles at the same lattice point, we temporarily turn off the pairwise force at the beginning of the simulation.  Of course, once nonlinear evolution begins we want particles to feel the pairwise force and so we turn the pairwise force back on at matter-radiation equality.  Because we use a pairwise force extended over $2$ extra grid cells and the CUBEP$^3$M fine force interpolation is via the nearest grid point method, if we set the particle separation to be $4$ grid cells then particles between neighboring lattice points also won't feel a fine force once they move out of their initial cell.  Thus, the number of particles we use is $6/8$ the regular value.  
    
\begin{figure}
            \includegraphics[width=0.45\textwidth]{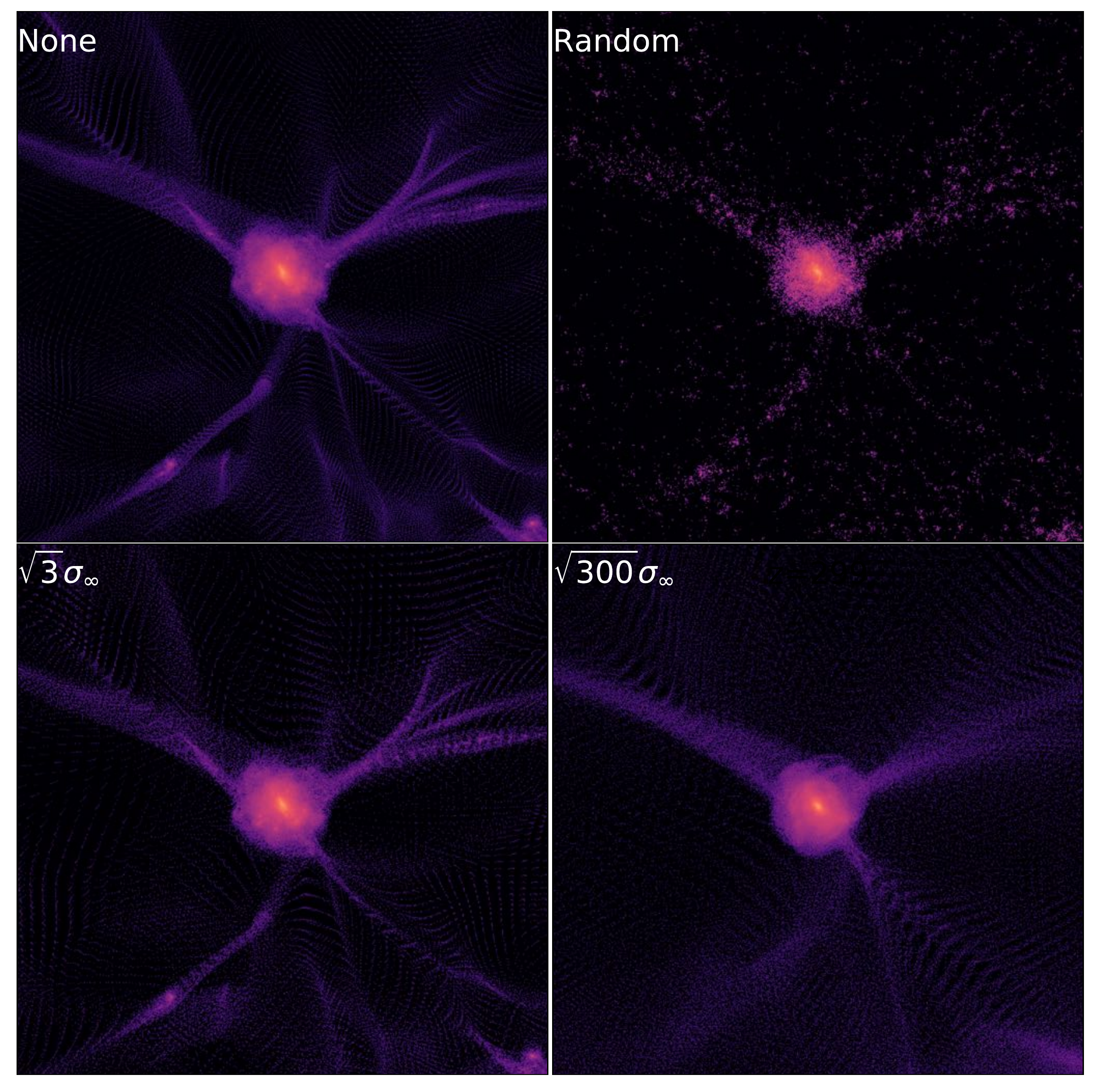}
            \caption{DAO density slice showing the effect of thermal velocities.  Top left panel is the same as in Fig.~\ref{fig:6slice}, top right panel is with completely random velocities assigned to each particle, and the bottom two panels use a regular velocity structure.}
            \label{fig:slice_u}
        \end{figure}

\begin{figure}
            \includegraphics[width=0.45\textwidth]{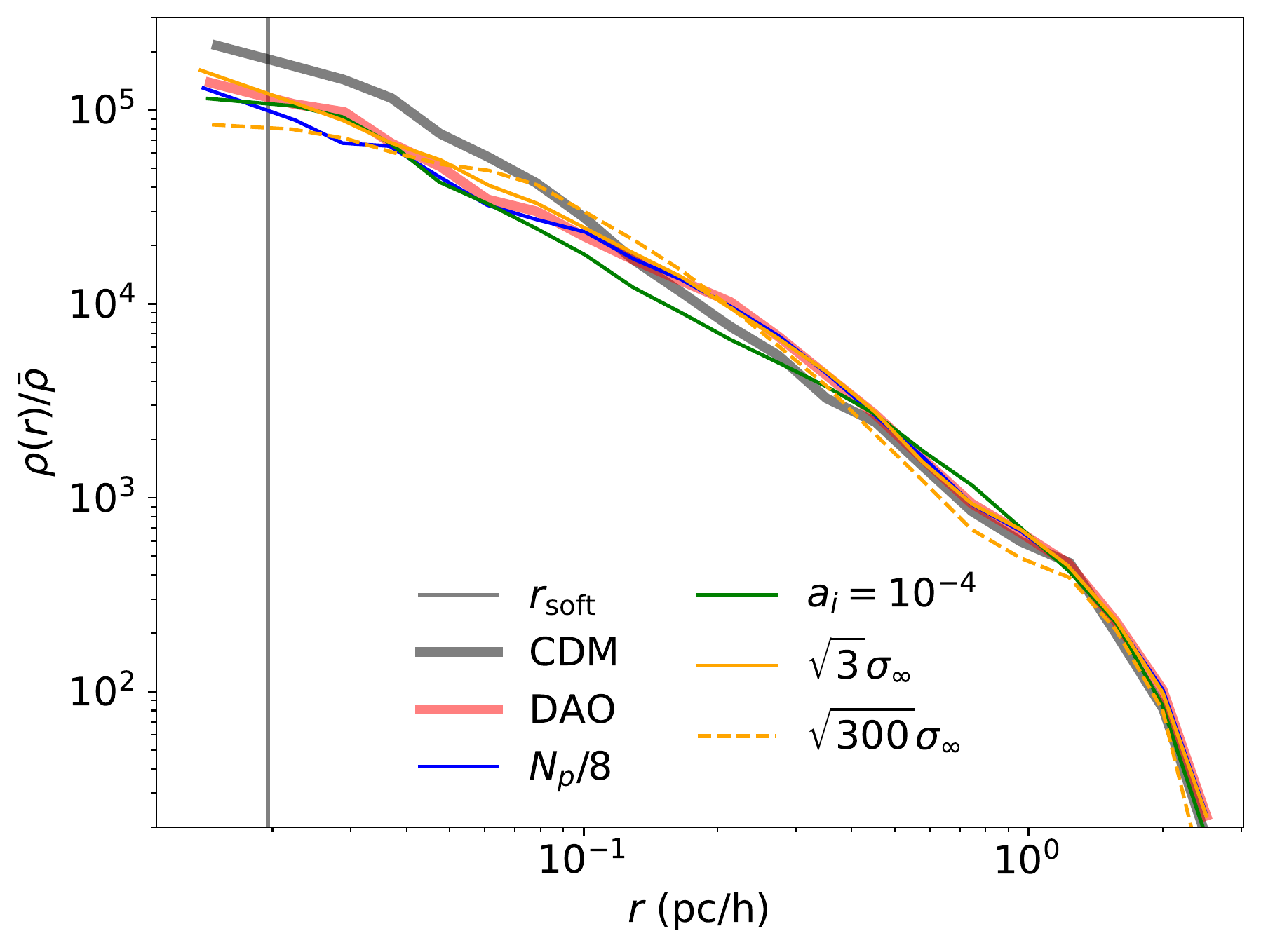}
            \caption{Convergence test of the largest halo profile with respect to the inclusion of thermal velocities.  The CDM and DAO profiles are the same as in Fig.~\ref{fig:rho}, whereas the two orange curves show profiles where thermal velocities have been included.  We also show a convergence test with respect to number of particles (blue) and initial redshift (green).}
            \label{fig:rho_u}
        \end{figure}

\begin{figure}
            \includegraphics[width=0.45\textwidth]{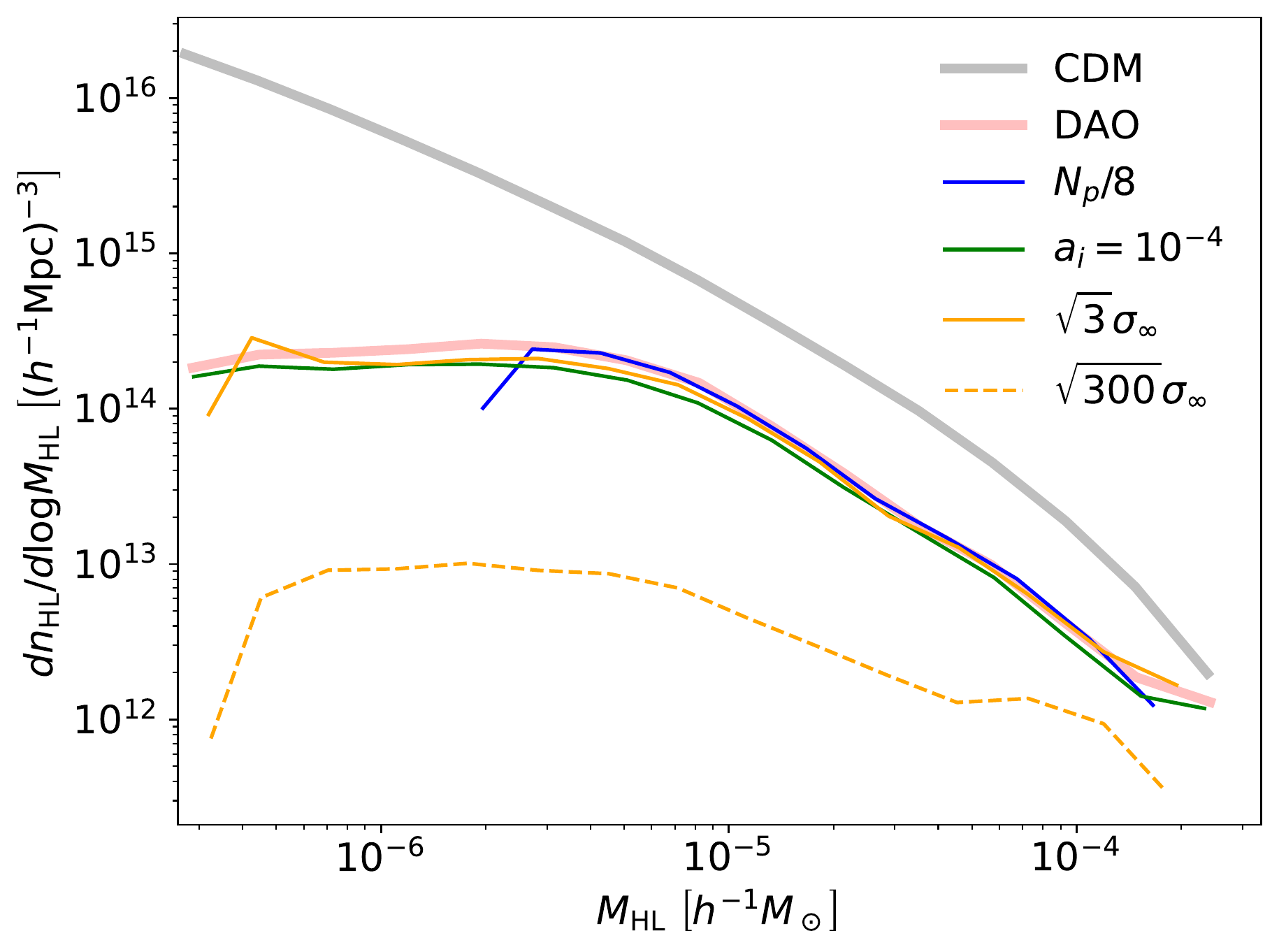}
            \caption{Convergence tests of the halo mass function at $z=299$.  The CDM, DPS, and DAO curves are the same as in Fig.~\ref{fig:hmf}.  The orange curves show the impact of adding thermal velocities, while blue and green curves show the effect of particle number and starting redshift.}
            \label{fig:hmf_u}
        \end{figure}

        We find that simulations run with this procedure (and for either value of $u$) do not have the artificial noise that occurs with pure random velocities.  We show in Fig.~\ref{fig:slice_u} the effects of various methods of including thermal velocities on the large halo shown in Fig.~\ref{fig:6slice}.  Using a regular velocity structure with $u=\sqrt{3}\sigma_\infty$ leads to a fairly similar result as the cold case, whereas using $u=\sqrt{300}\sigma_\infty$ smooths out some of the filamentary structure.  The random thermal velocities have additional fragmentation due to Poisson noise and we therefore do not consider it further.  We show in Fig.~\ref{fig:rho_u} the density profile of this halo with the random motions.  We find that the $\sqrt{3}\sigma_\infty$ simulation has essentially the same profile as without random motions,  whereas $\sqrt{300}\sigma_\infty$ is only changed a little.  We furthermore have performed a test where we just halve the number of particles and find excellent overall convergence.  The halo mass function, shown in Fig.~\ref{fig:hmf_u} is mostly unchanged for the lower value of $u$, however we find that the mass function as a whole is substantially lower for the larger $\sqrt{300}\sigma_\infty$.  We therefore conclude that our results should be relatively robust to thermal motions, but some changes could occur since some particles will have larger than average velocities.  It furthermore appears that introducing regularity in velocity space is a promising way to include thermal effects in cold dark matter as well.
    
\begin{figure}
            \includegraphics[width=0.45\textwidth]{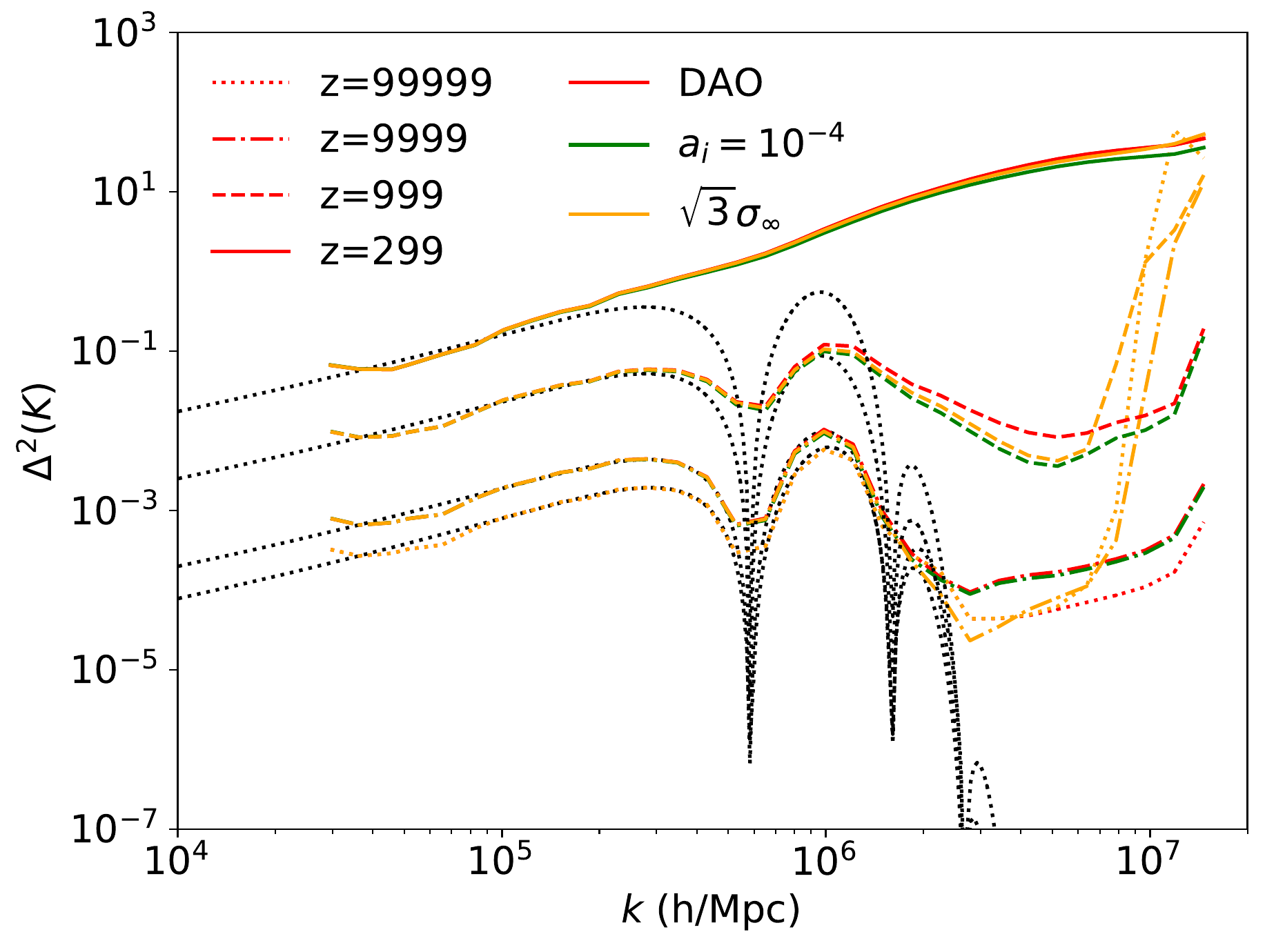}
            \caption{Convergence test of the DAO power spectrum at various redshifts.  Orange curves show the effect of thermal velocities, while green curves show the effect of starting redshift.}
            \label{fig:powerSz}
        \end{figure}

        We now consider how thermal velocities interact with the starting redshift of our simulations.  In Fig.~\ref{fig:powerSz} we show the power spectrum at $a=10^{-5},10^{-4},10^{-3}$, and $10^{-2.5}$ for the DAO simulation without thermal velocities (red) and with thermal velocities (orange).  In green we show a simulation without thermal velocities, but starting at $a_i=10^{-4}$ instead of $a_i=10^{-5}$.  We note that there is always some numerical noise floor in the initial conditions rather than a pure cutoff.  Furthermore, from $a=10^{-5}$ to $a=10^{-4}$ this floor grows without thermal velocities but is suppressed when they are included.  At $a=10^{-3}$ and $z=299$ we find more substantial deviations between the $a_i=10^{-5}$ simulation and the $a_i=10^{-4}$ one.  The thermal velocities appear to suppress the power spectrum at $a=10^{-3}$ leading to better agreement with a later start; however, by $z=299$ the situation is reversed and the orange curve agrees better with the simulation started earlier.  This is somewhat curious and further motivates a more complete treatment of thermal velocities.  For now, we consider this a source of error in our results.  The boost factor differs by $\sim25\%$ between $a_i=10^{-5}$ and $a_i=10^{-4}$ simulations at both $z=999$ and $z=299$.  We also find suppression in the halo mass function, shown in Fig.~\ref{fig:hmf_u}, while the halo profile in Fig.~\ref{fig:rho_u} is more robust.

    \subsection{Artificial Fragmentation}
    \label{app:con:af}
    
\begin{figure}
            \includegraphics[width=0.45\textwidth]{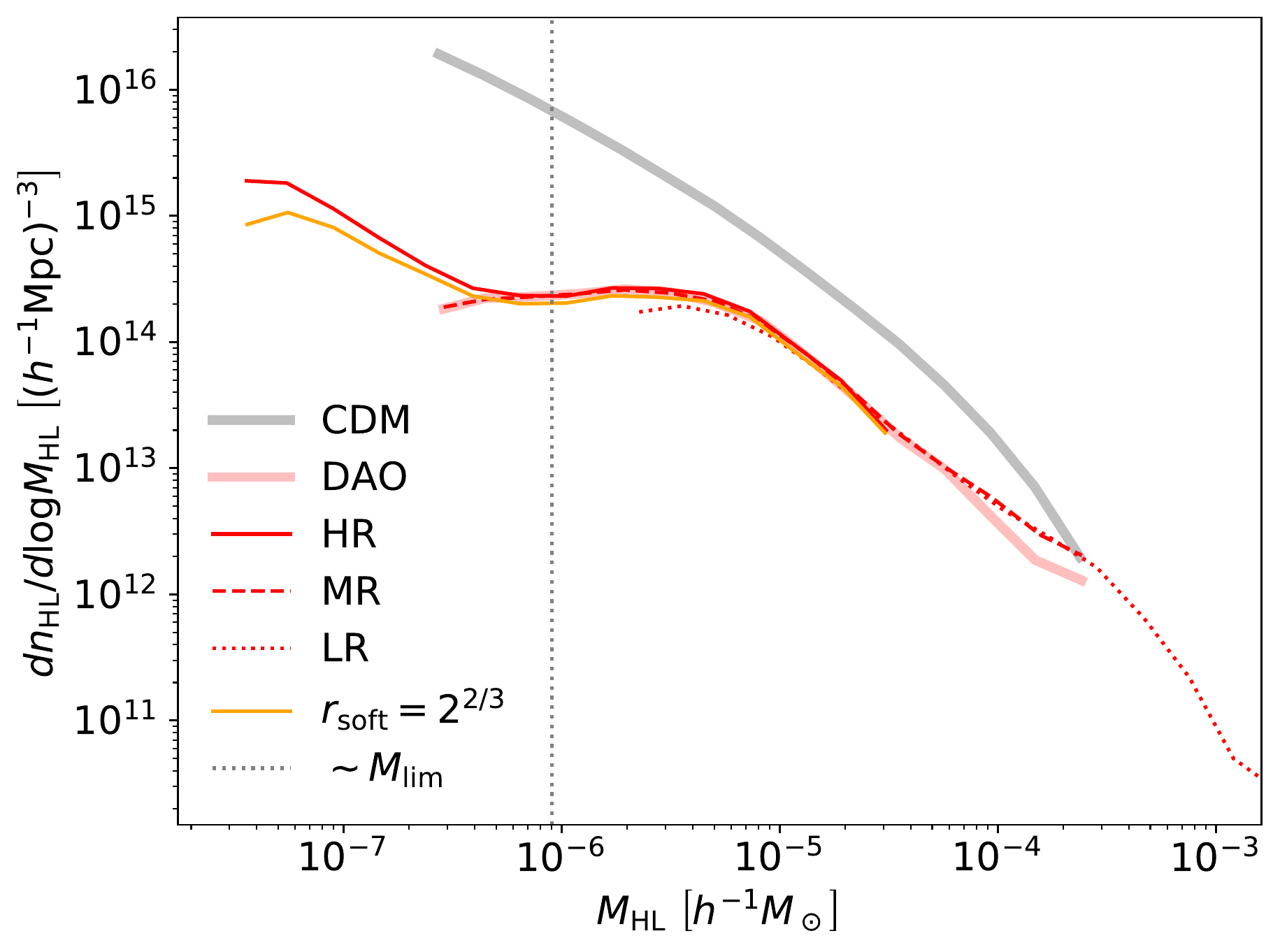}
            \caption{Convergence tests of the halo mass function at $z=299$ with respect to resolution.  The CDM and DAO bands are the same as in Fig.~\ref{fig:hmf} whereas the red curves show the effect of changing resolution and the orange curve the effect of reducing the force softening length.}
            \label{fig:hmf_af}
        \end{figure}

        To test the amount of artificial halos in the halo mass function, we start by performing standard convergence tests with respect to particle number and length scale.  We increase the number of particles from $2\times768^3$ to $2\times1024^3$ and run simulations in volumes of $(200{\rm\ h^{-1}kpc})^3$, $(400{\rm\ h^{-1}kpc})^3$ and $(800{\rm\ h^{-1}kpc})^3$ labelled HR, MR, and LR, and with the MR simulation having an equivalent resolution to that of the main simulations.  We show the results in Fig.~\ref{fig:hmf_af}.  We find that on the scales probed there is good agreement.  However, in the HR simulation we observe a substantial uptick in halos around $M_{\rm lim}\sim9\times10^{-7}M_\odot$, corresponding to a value of $k_p\sim5\times10^{5}{\rm\ h^{-1}kpc}$, which is broadly consistent with the DAO transfer function.  In the LR simulation we find that heavy halos are not quite as suppressed as appears in the main simulation, which could be due to cosmic variance of the simulations.

\begin{figure}
            \includegraphics[width=0.45\textwidth]{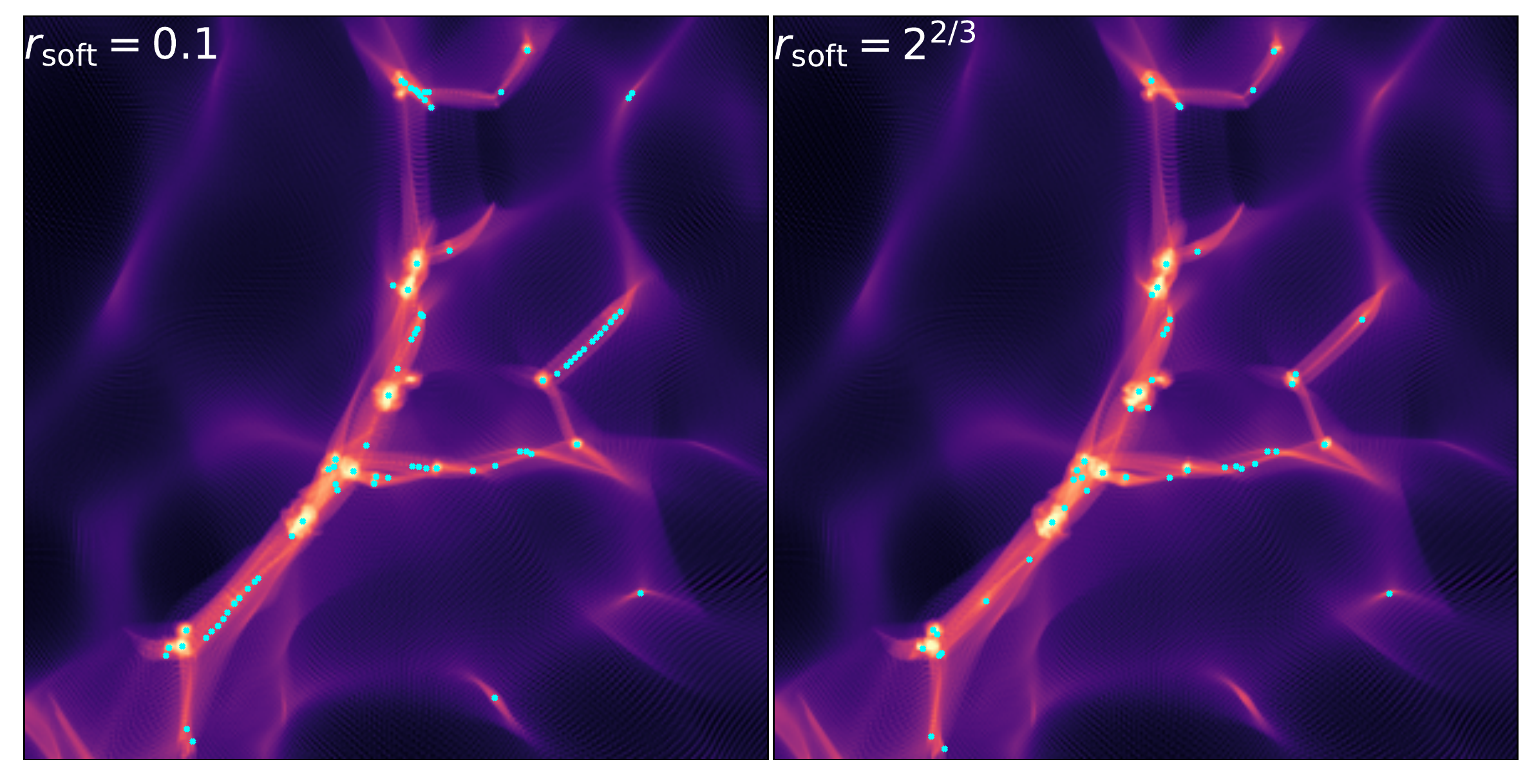}
            \caption{A zoomed in DAO density field enclosing a $30\times30\times4{\rm\ (h^{-1}kpc)^3}$ volume comparing the effects of the standard force resolution (left) and one where the softening length is set to the inter-particle separation (right).  Halos are shown as cyan crosses; artificial fragmentation is reduced with larger force softening.}
            \label{fig:af}
        \end{figure}

        Since it has been suggested that artificial halos arise from mismatched mass and force resolution \citep{bib:Melott1989,bib:Angulo2013} we have also tested running the HR simulation with substantially reduced force resolution.  To do this, we reran the HR simulation but set the softening length to be the inter-particle spacing $L/N_p^{1/3}$ ($2^{2/3}$ fine grid cells).  We find that some of the artificial halos along filaments are indeed removed.  We show an illustrative region of the simulation in Fig.~\ref{fig:af} where the characteristic inter-spaced halos along filaments are not found with reduced force resolution.  However, we show in Fig.~\ref{fig:hmf_af} that there is still a substantial uptick in the halo mass function.  While a small uptick is also observed in the particle-mesh simulations of \citet{bib:Angulo2013}, the one we find appears much more substantial. This could be due to a number of things including a lack of convergence, residual noise in the force calculation, the halofinder finding non/proto-halo structures \citep{bib:Angulo2013}, or potentially some quirk of our DAO initial conditions.  Understanding these very lightest halos is certainly important, and will require a more detailed investigation.
        
    \subsection{Baryonic Collapse}
    \label{app:con:bc}
    
        To test the potential consequences of baryonic clustering on our larger volume simulations, we run test simulations where we assume baryons cluster exactly like dark matter after a certain redshift.  We set the initial conditions after recombination using just the CDM transfer function and increase $\Omega_c\rightarrow \Omega_c+\Omega_b$.  We use an initial seed corresponding to \#3 in Table~\ref{tab:halomass}.  When we set the initial redshift to be immediately after recombination, $z_i=999$, we find a large enhancement as the number of halos with masses greater than $10^5 {\rm\ h}^{-1}{\rm M}_\odot$ is 64, and the heaviest halo is $1.4\times10^6 {\rm\ h}^{-1}{\rm M}_\odot$.  However, if we use $z_i=199$ then only 18 halos have masses $10^5 {\rm\ h}^{-1}{\rm M}_\odot$ and the heaviest halo has mass $6\times10^{5}{\rm\ h}^{-1}{\rm M}_\odot$.  This analysis is not meant to be quantitative, but rather illustrative of the potential role baryons may play if they catch up to CDM earlier than expected due to an enhanced power spectrum (as is the case in some halos studied in \citet{bib:Hirano2015}).  A correct understanding of baryonic effects will require hydrodynamic simulations, but we can expect our results to underestimate the true amount of clustering.  

\end{document}